\documentclass[a4paper,10pt]{article}
\usepackage[latin1]{inputenc}
\usepackage{float}
\usepackage[greek, frenchb, english]{babel}
\usepackage{amsmath}
\usepackage{amsfonts}
\usepackage{graphicx}
\usepackage{epstopdf}
\usepackage{subfigure}
\usepackage{array}
\usepackage{color}
\usepackage{setspace}
\usepackage{amssymb,mathrsfs}
\usepackage{bm}
\usepackage{cite}
\usepackage{authblk}

\graphicspath{{./figures/}}

\begin{document}

\title{Physics of a toy geyser}

\author{Martin Brandendourger}
\author{St\'ephane Dorbolo}
\author{Baptiste Darbois Texier}

\affil{\small{GRASP}, Physics Dept., ULg, B-4000 Li\`{e}ge, Belgium.}

\date{}     

%\keywords{}
%\classification{}

\maketitle

\begin{abstract}
A geyser can be reproduced by a toy experiment composed of a water pool located above a water reservoir, the two being connected by a long and narrow tube. When the bottom reservoir is heated, the system experiences periodic eruptions of hot water and steam at the top similarly to the geyser effect occurring in nature. The eruption frequency of a toy geyser is inspected experimentally as a function of the heating power and the height of the set-up. We propose a thermodynamic model of this system which predicts the time between eruptions. Besides, the geometric and thermodynamic conditions required to observe a geyser effect are discussed. The study of the toy geyser is then extended to the case of two reservoirs connected to the same tube. In such a configuration, an eruption in a reservoir may entrain the eruption of the other. Such a coupled system adopts a complex time-evolution which reflects the dynamics of natural geysers. We analyze the behavior of a toy geyser with two reservoirs by the way of statistical tools and develop a theoretical model in order to rationalize our observations.
\end{abstract}  

\section*{Introduction}

\begin{figure}[h!]
\centering
	\begin{minipage}[c]{0.30\textwidth}
  		\centering
  		\hspace{0.2cm}(a)\\
		\vspace{0.2cm}
		\includegraphics[height=5cm]{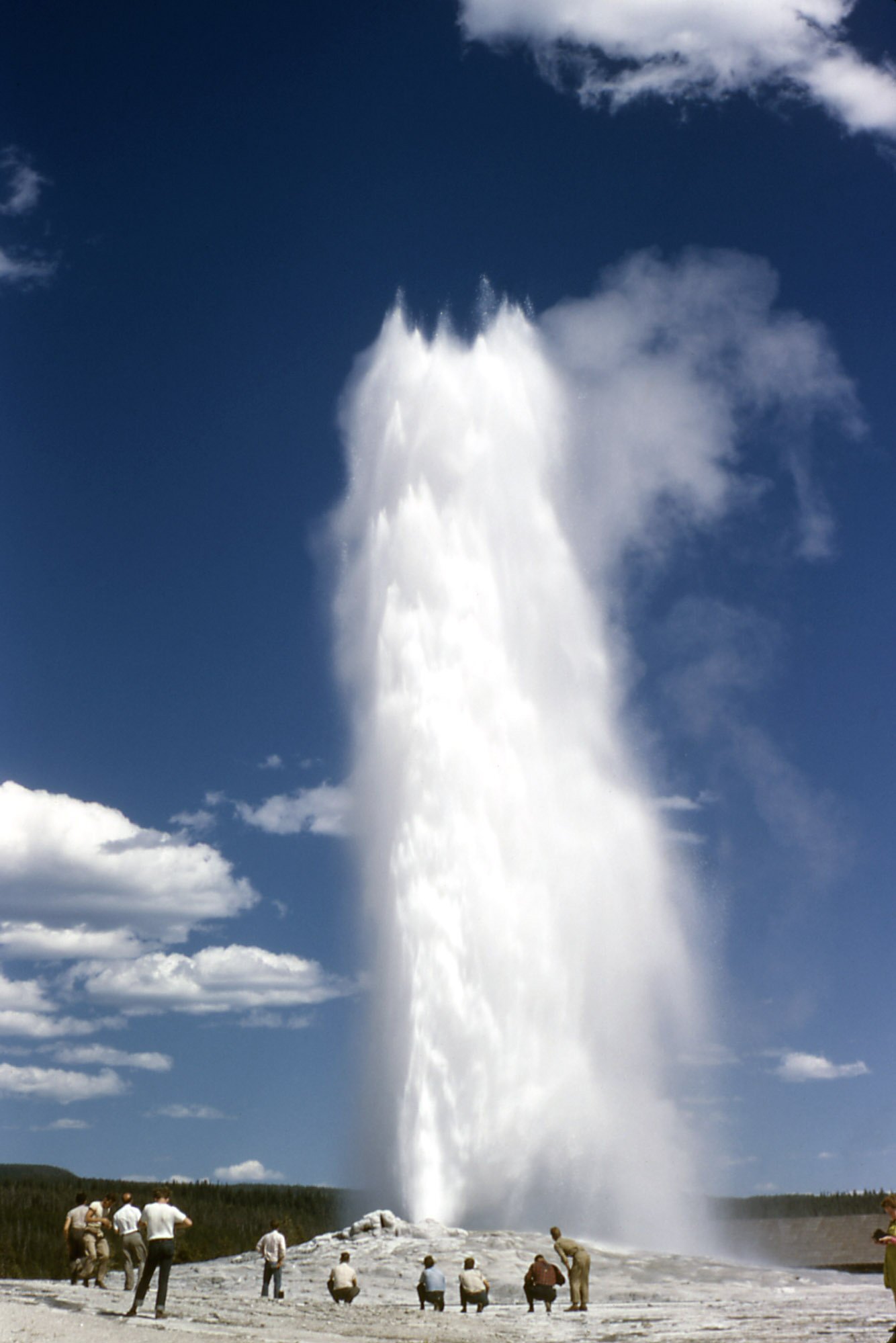}
 	\end{minipage}%
		\begin{minipage}[c]{0.30\textwidth}
  		\centering
  		\hspace{0.1cm}(b)\\
		\vspace{0.2cm}
		\includegraphics[height=5cm]{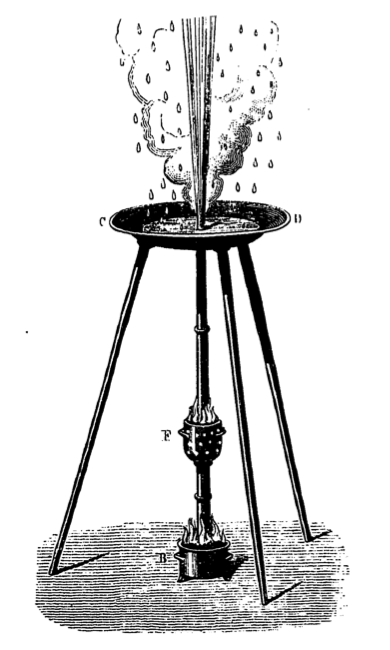}
 	\end{minipage}%
	\begin{minipage}[c]{0.30\textwidth}
	\centering
	\hspace{0.2cm}(c)\\
	\vspace{0.2cm}
	\includegraphics[height=5cm]{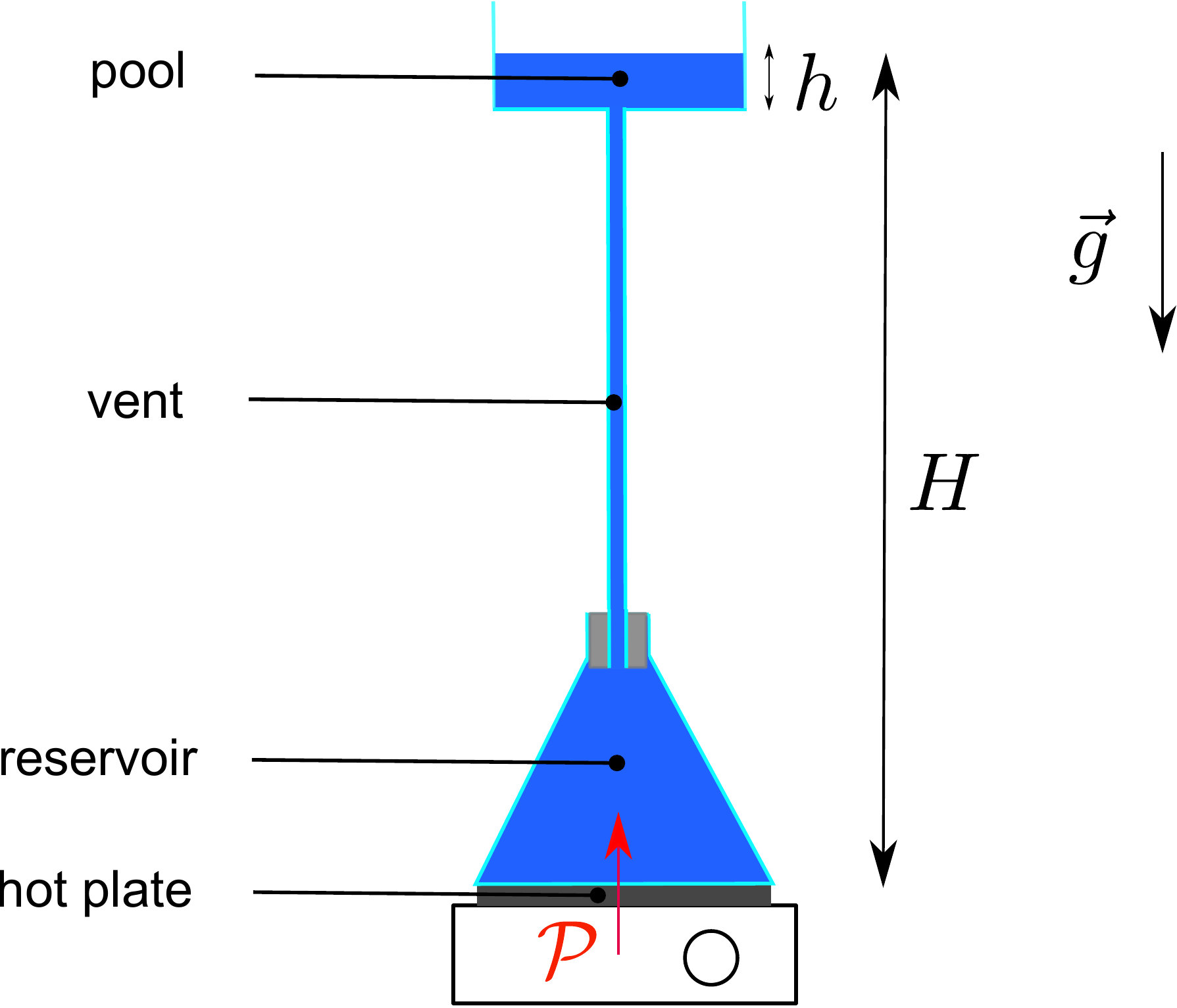}	
	\end{minipage}
 	\caption{(a) People watching Old Faithful erupt from geyser cone, Yellowstone National Park, 1948. (b) Toy model of a geyser by Tyndall \cite{tyndall1881heat}. (c) Sketch of the experimental set-up. A glass reservoir containing a mass $M$ of water at a temperature $T_r$ was connected to a pool by a long vent of radius $R$. The water in the pool was maintained at a constant temperature $T_p$. The system was completely filled with a column of water of height $H$ chosen between 1.6 and 11.5 m and the pool has a depth $h$. A heating plate provided a specific heat power $\mathcal{P}$ to the water of the reservoir.}
		\label{fig:geyser}
\end{figure}

A geyser can be defined as a periodically eruptive spring resulting from Earth's thermal activity. One of the most famous geyser is the Old Faithful [Fig. \ref{fig:geyser}(a)] in the Yellowstone National park in the United States. It erupts about every 90 minutes and projects hot water up to 56 m high\cite{rinehart1969old,azzalini1990look}. Probably the first scientist to describe the principle of a geyser was Robert Bunsen in 1845 after the Danish government sponsored him to study the eruption of Mount Hekla in Iceland \cite{bunsen1904gesammelte}. He lowered a thermometer into geyser depths and discovered that they had at their base a reservoir of \og superheated \fg water. He understood that when the water in the reservoir boils, the liquid in the vent is ejected upwards and produces an eruption of hot water and steam. Bunsen and his student Tyndall built a toy geyser based on the same principle as illustrated in Fig. \ref{fig:geyser}(b)  \cite{tyndall1865heat}. Such a demonstrative experiment showed eruptions regularly spaced in time that confirmed their intuitions on the fundamental principles underlying natural geysers. The realization of a geyser at reduced scale contributed to promote the Bunsen's theory for geyser among the geophysical community \cite{dowden1991dynamics,saptadji1995modelling}. Nowadays, similar set-ups are present in various Science Museum like the Exploratorium in San Francisco (USA) and Source-o-Rama in Chaudfontaine (BE) and are also currently used for research purpose in order to understand mechanisms underlying a geyser eruption \cite{toramaru2013mass,adelstein2014geyser}.

In nature, the dynamics of a geyser can be much more complex time series that the one of a toy geyser as revealed by \textit{in situ} data collection in Yellowstone USA \cite{rinehart1969old,azzalini1990look,hutchinson1997situ} or at El Tatio in Chile \cite{namiki2014cobreloa,munoz2015dynamics}. Such a complexity can be attributed to a high sensitivity of geysers dynamics towards water and steam inflow \cite{munoz2015dynamics}, barometric pressure, tidal forces, tectonic stresses \cite{rinehart1972fluctuations} or reservoirs interconnection \cite{rojstaczer2003variability}. Over the past decades, crucial steps in the theoretical modelling of geyser eruptions have been achieved \cite{dowden1991dynamics,rudolph2012mechanics,o2013model} but an extensive description of the problem is still to uncover.

%Historically, the word geyser comes from \textit{Geysir}, the name of an erupting spring at Haukadalur (Iceland) which, in turn, comes from the Icelandic verb \textit{geysa}, \og to gush \fg.

%The questions we would like to address in the article are: \textit{Can we predict the time interval between eruptions for a model geyser? In what cases no eruptions occur? How similar is this phenomenon in nature?}

Throughout this paper we aim at studying and predicting the dynamics of a toy geyser and understanding its similarities with natural geysers. First, we reproduced a model geyser similar to the one of Tyndall. The erupting dynamics of such a geyser has been inspected as a function of the injected heat power and its geometrical characteristics. The different stages of a geyser life and the evolution of the reservoir temperature over time are described in Section \ref{sec:experiments}. Afterwards, we develop in Section \ref{sec:model} a theoretical model based on thermodynamic notions and fluid dynamics in order to rationalize our observations. Besides, we observed that the geyser effect can vanish because of intense vaporization or thermal convection in the vent. The required conditions to produce a geyser effect are inspected in Section \ref{sec:discussion} and compared with conditions occurring in nature. Finally, we showed that the complex dynamics of natural geysers can be approached by interconnecting two reservoirs to a single vent as detailed in Section \ref{sec:two_reservoirs}.

%A glass reservoir containing a mass $M$=1.2 kg of water at a temperature $T_r$ is connected to a pool by a long vent of radius $R=$3 mm. The water in the pool is maintained at a constant temperature $T_p=30^\circ \rm{C}$. The system is completely filled with a column of water of height $H=$ 1.6 and 11.5 m and the pool has a depth $h=$0.04 m. A heating plate provides a specific heat power $\mathcal{P}$ to the water of the reservoir which varies between 30 and 220 W/kg.

\section{Experiments}\label{sec:experiments}

\subsection{Primary observations}

\begin{figure}[h!]
\centering
\begin{minipage}[c]{0.50\textwidth}
\centering
\hspace{0.8cm}(a)\\
\vspace{0.2cm}
\includegraphics[height=5cm]{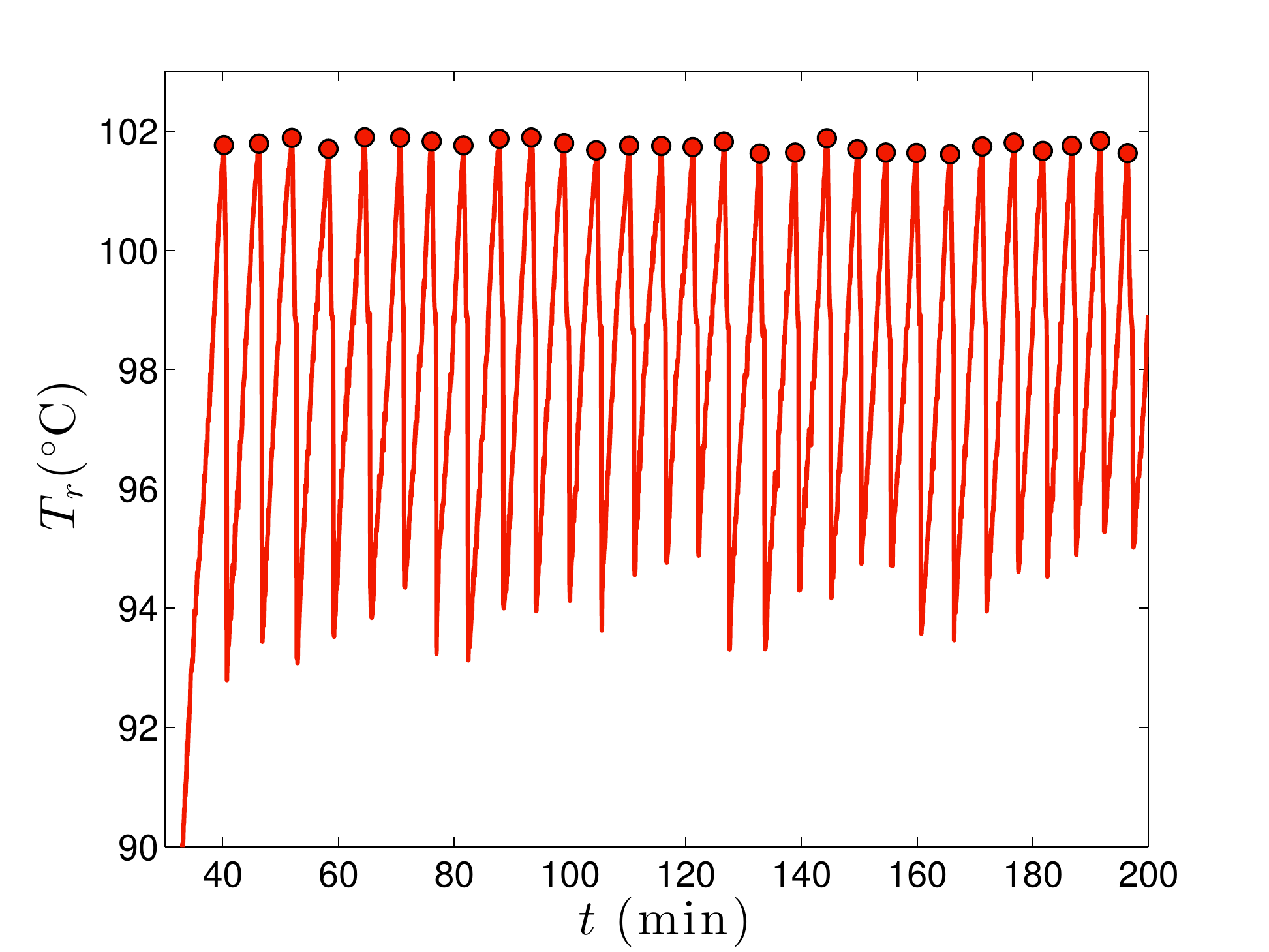}
\end{minipage}
\begin{minipage}[c]{0.30\textwidth}
\centering
\hspace{0.4cm}(b)\\
\vspace{0.2cm}
\includegraphics[height=5cm]{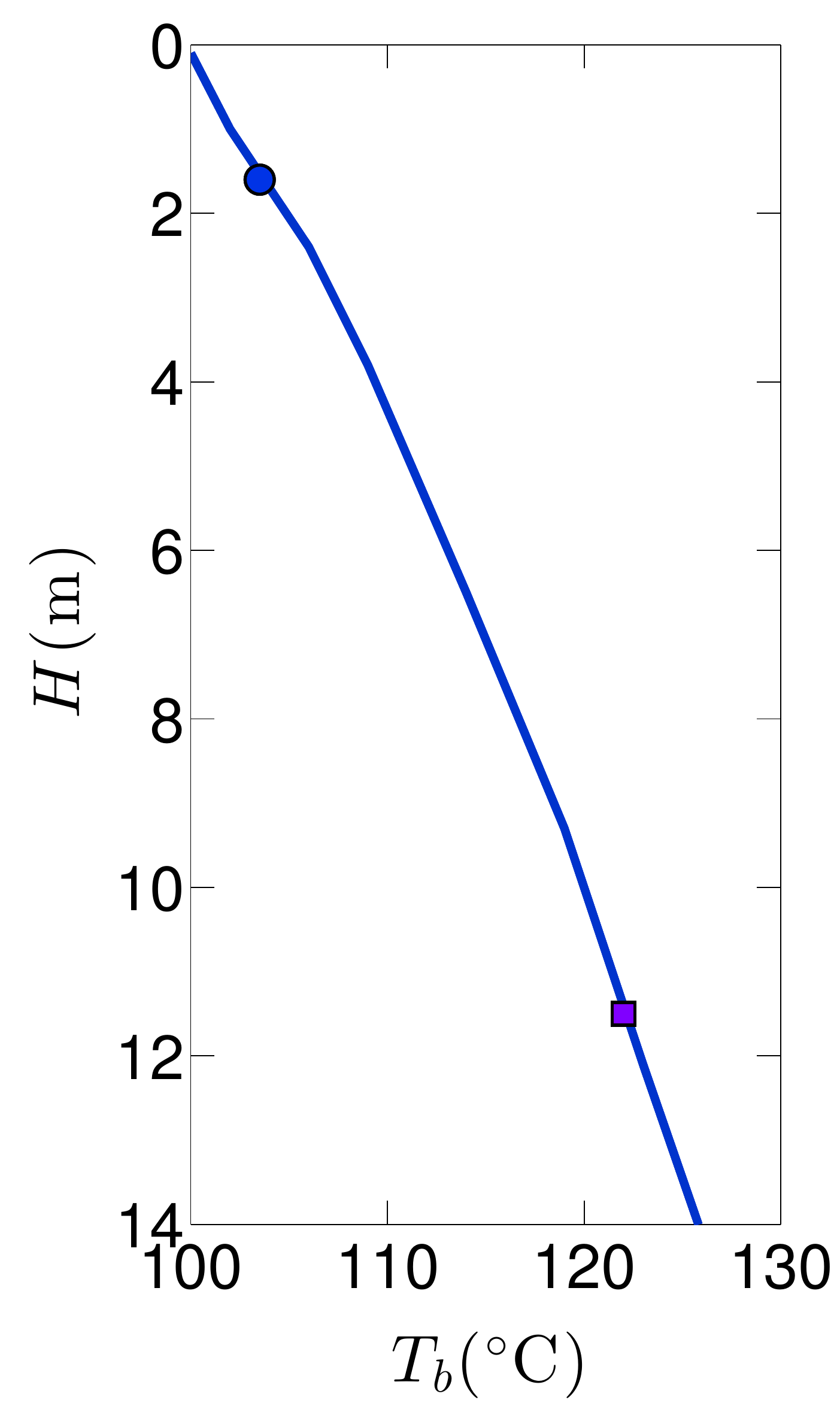}
\end{minipage}
\caption{(a) Evolution of the temperature $T_r$ in the reservoir over time for a toy geyser of height $H=1.60$ m, specific heating power $\mathcal{P}=130$ W/kg, pool height $h= 4.0$ cm and vent radius $R=3.0$ mm. Dots indicate the time of eruptions. (b) Boiling temperature $T_b$ of water as a function of the height $H$ of the overlying column extracted from \cite{lide2004crc}. Symbols show the parameters of the experiments presented in this study.}
\label{fig:t_T_r}
\end{figure}

\begin{figure}[h!]
\centering
\includegraphics[height=14cm]{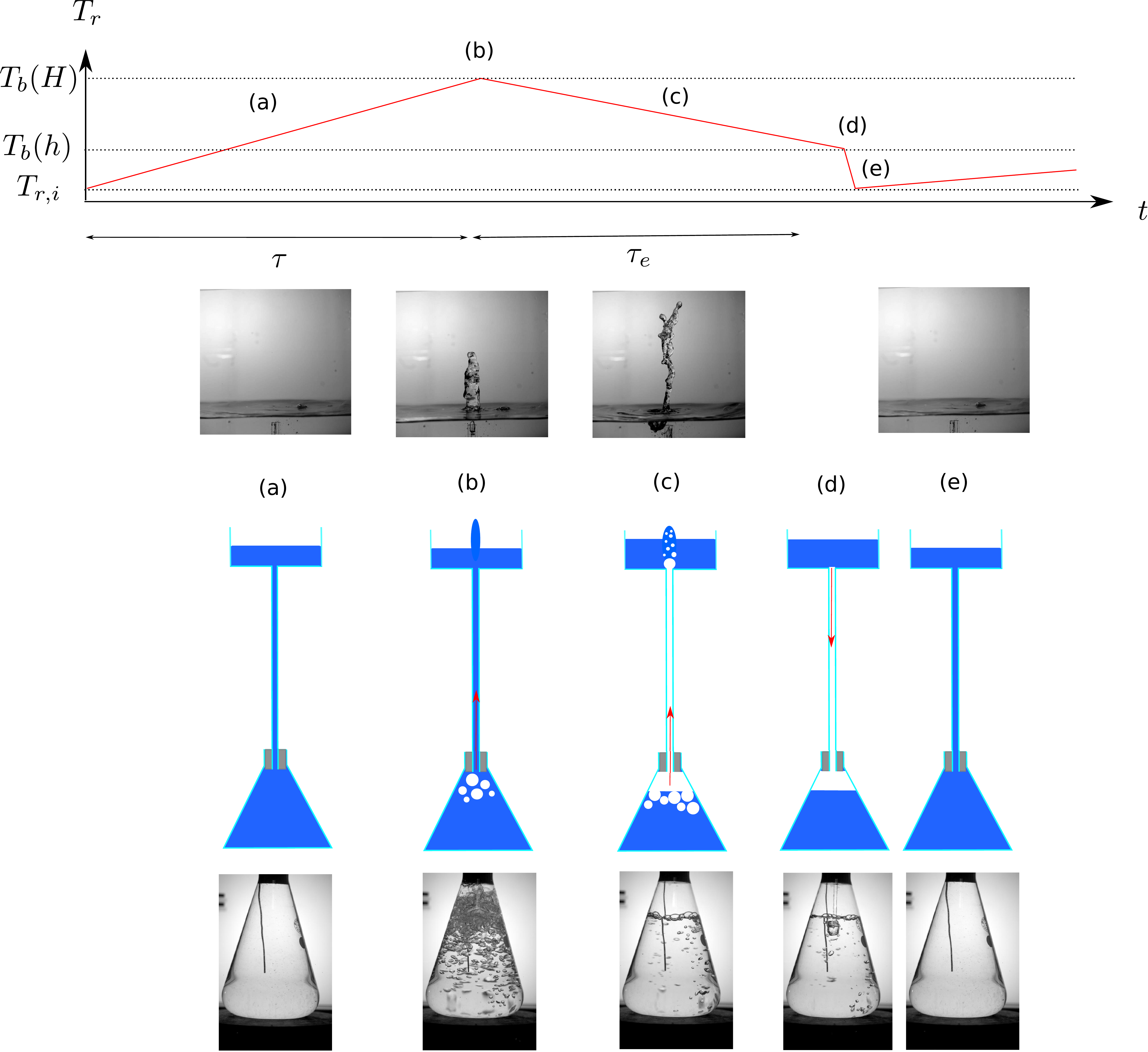}
\caption{Sketch showing the different stages of the life of a toy geyser: (a) Heating of the reservoir up to the boiling temperature $T_b (H)$; (b) Water boiling in the reservoir provokes an eruption; (c) Vapor emission throughout the vent; (d) End of the eruption, the water from the pool refills the reservoir; (e) The system returns to its initial state. The typical evolution of the reservoir temperature $T_r$ over time is shown by the way of the red solid line in the graphic at the top. The pool temperature is maintained at $T_p=30^\circ$C. Bottom and top pictures correspond to the reservoir and the pool at different times of the geyser cycle.}
\label{fig:steps}
\end{figure}

A toy geyser was realized with a glass reservoir connected to a pool through a long and narrow vertical vent as sketched in Fig. \ref{fig:geyser}(c). The system was completely filled with water. The total height of the liquid column from the bottom of the reservoir to the liquid surface is denoted $H$. The depth of the pool was $h$ and the vent radius $R$. The reservoir, which contained a mass of water $M$, was heated from below by a hot plate maintained at a constant temperature. Thus, the water of the reservoir received a specific heat power $\mathcal{P}$. When heated continually, the toy geyser undergoes periodic erupting events spaced between quiet phases. This periodic behavior is illustrated by the evolution of the temperature $T_r$ measured in the middle of the reservoir as shown in Fig. \ref{fig:t_T_r}(a) for a geyser height $H=1.60$ m and a specific heating power $\mathcal{P}=130$ W/kg. After a first heating phase, the temperature in the reservoir follows a saw-tooth evolution between about 93$\,^\circ$C and 102$\,^\circ$C. A single period of the temperature of the reservoir is reported typically in Fig. \ref{fig:steps} and consists of several steps. First, the water in the reservoir is heated up to the boiling point $T_b(H)$ at a pressure corresponding to the water column of height $H$ [stage (a) in Fig. \ref{fig:steps}]. The boiling temperature of water as a function of the height of the overlying column is recalled in Fig \ref{fig:t_T_r}(b) and is equal to 102$\,^\circ$C for $H=1.60$ m. When the water in the reservoir reaches the corresponding boiling point, an intense production of vapor bubbles occurs. Buoyancy induces the upward motion of vapor bubbles in the vent which entrains the water above and produces an eruption of hot water and steam at the top of the toy geyser [stage (b) in Fig. \ref{fig:steps}]. After the ejection of the liquid in the vent, the water in the reservoir keeps on boiling and the vapor continues to escape throughout the vent toward the pool [stage (c) in Fig. \ref{fig:steps}]. The total amount of water transferred from the reservoir to the pool at this stage is denoted $\Delta M$. As the boiling goes on, the vapor production cools down the water remaining in the reservoir. When this latter reaches the boiling temperature $T_b(h)$ corresponding to the pressure of a water column of height $h$, the vapor emission decreases till the vapor flow can not sustain the water in the pool which finally falls back into the reservoir [stage (d) in Fig. \ref{fig:steps}]. At this moment, the eruption phase is finished. The eruption has last a time $\tau_e$. The temperature in the reservoir reaches $T_{r,i}$, which is the equilibrium temperature resulting from the mixture of a mass of water $M-\Delta M$ at the temperature $T_b (h)$ and a mass of water $\Delta M$ coming from the pool maintained at a constant temperature $T_p$ [stage (e) in Fig. \ref{fig:steps}]. After that, the temperature in the reservoir increases from $T_{r,i}$ to $T_b(H)$ in a time $\tau$ before an other eruption starts. Meanwhile, the temperature of the pool is thermalized to $T_p=30^\circ$C throughout the duration of the experiment.

\subsection{Measurements}

\begin{figure}[h!]
\centering
	\begin{minipage}[c]{0.5\textwidth}
  		\centering
  		\hspace{0.5cm}(a)\\
		\vspace{0.2cm}
		\includegraphics[width=7cm]{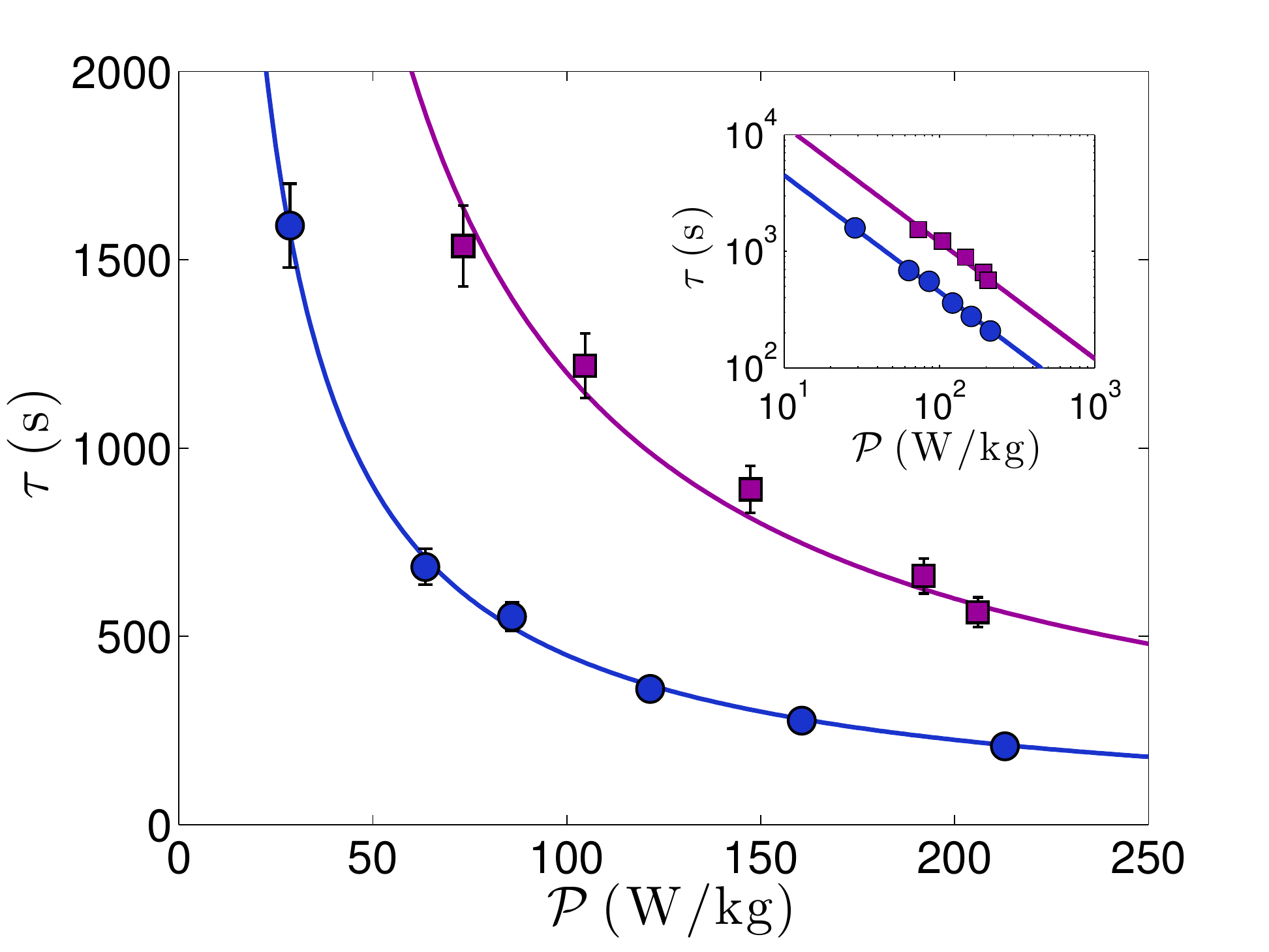}
 	\end{minipage}%
	\begin{minipage}[c]{0.5\textwidth}
  		\centering
  		\hspace{0.5cm}(b)\\
		\vspace{0.2cm}
		\includegraphics[width=7cm]{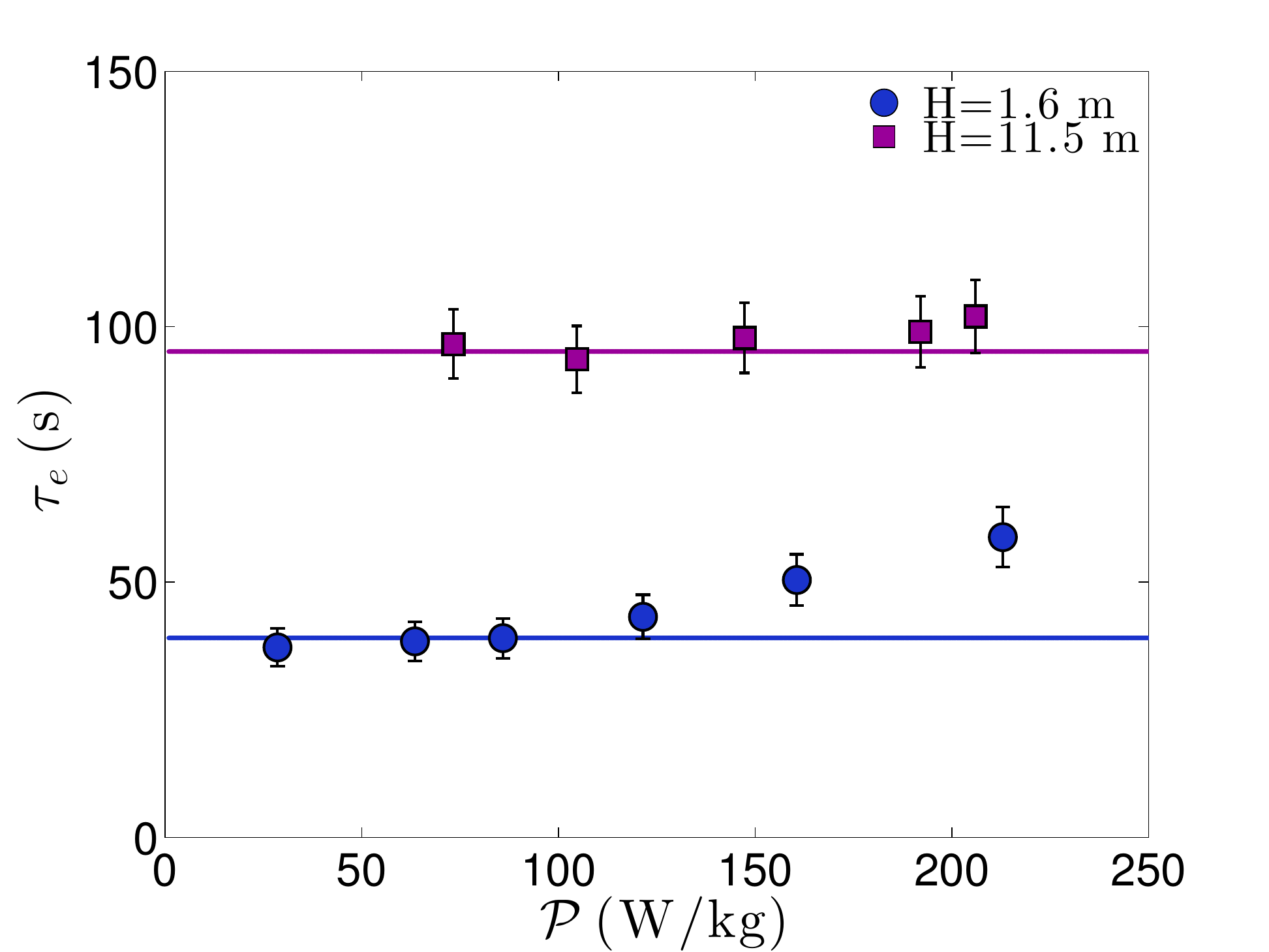}
 	\end{minipage}%
 	\caption{Mean values of the time between geyser eruptions $\tau$ as a function of the specific heat power $\mathcal{P}$ for different heights of the water column $H$. The blue dots and purple squares correspond to $H=1.6$ m and $H=11.5$ m. The inset presents the data with logarithmic scales. Solid lines show the predictions of Eq. (\ref{eq:geyser_period}). (b) Mean erupting time $\tau_e$ as a function of the specific heat power $\mathcal{P}$ for different heights of the water column $H$. The blue dots and purple squares correspond to $H=1.6$ m and $H=11.5$ m. Solid lines show the predictions of Eq. (\ref{eq:geyser_tau_e}).}
		\label{fig:experimental_results}
\end{figure}

\begin{figure}[h!]
\centering
\includegraphics[width=7cm]{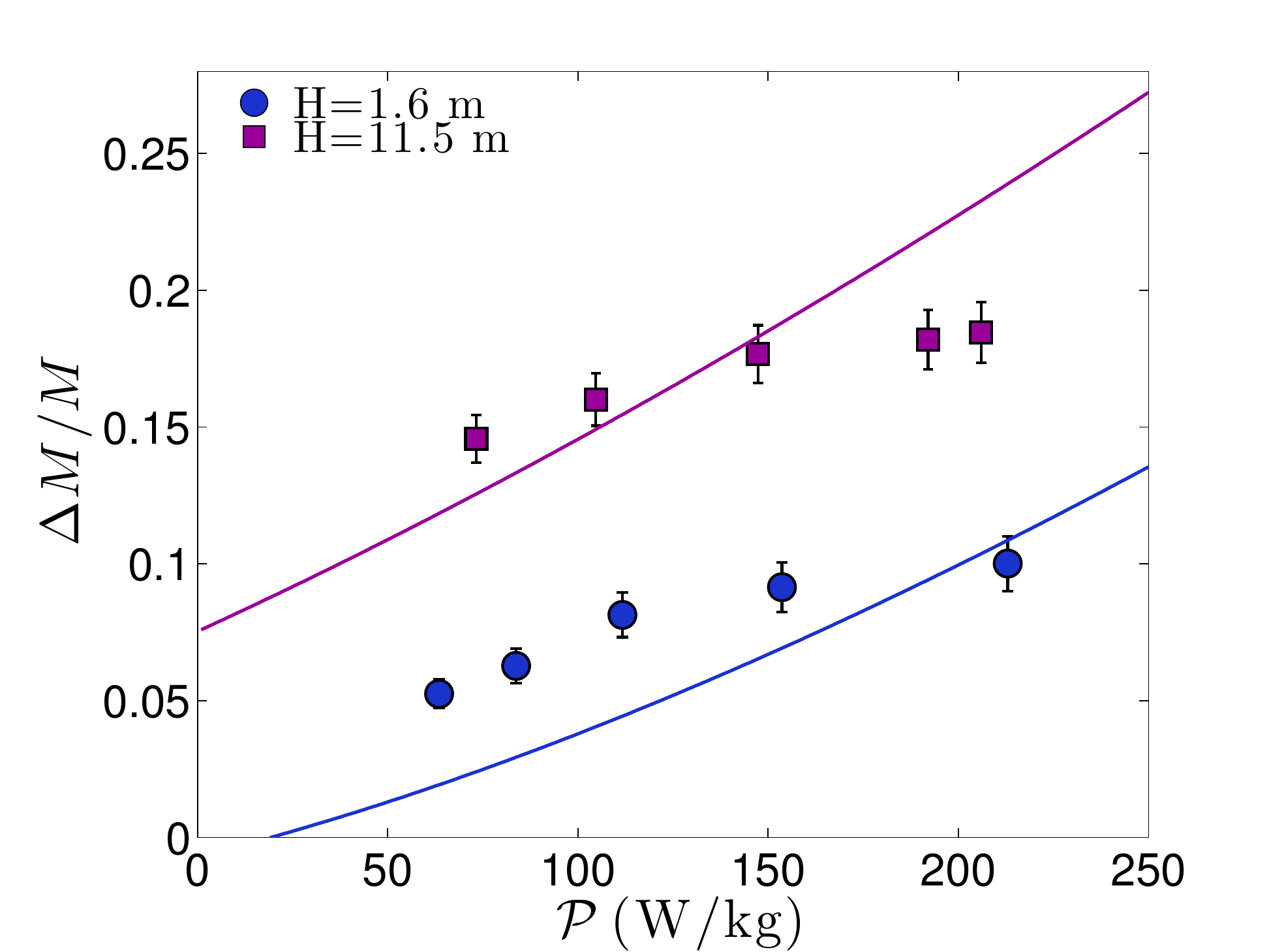}
\caption{Mean mass fraction $\Delta M / M$ of the reservoir exchanged with the pool during an eruption as a function of the specific heating power $\mathcal{P}$ for different water column heights $H$. The blue dots and purple squares correspond respectively to $H=1.60$ m and $H=11.50$ m. Solid lines show the predictions of Eq. (\ref{eq:mass_exchanged}).}
		\label{fig:experimental_results_dmm}
\end{figure}

The two characteristics time $\tau$, $\tau_e$ and the exchanged mass $\Delta M$ were measured as a function of the parameters of the experiments: the specific heating power $\mathcal{P}$ and the geyser height $H$. The specific heating power received by the water in the reservoir is estimated during the first heating step  throughout the relation

\begin{equation}
\mathcal{P}= \, c \, \frac{dT_r}{dt}
\label{eq:Clapeyron}
\end{equation}

\noindent where $c$ is the specific thermal capacity of water ($c=4.18\, \rm{kJ.K^{-1}.kg^{-1}}$). One remarks that $\mathcal{P}$ takes into account the heat transfer lost by the reservoir toward its environment. In our experiments the specific heating power could be varied between 30 and 220 W/kg. Also, we inspected two different geyser heights in this study, $H=1.60$ m and $H=11.50$ m, for which the boiling temperatures of water are respectively $102 \,^\circ \rm{C}$ and $120 \,^\circ \rm{C}$ for a standard outside atmospheric pressure. The mean values of $\tau$, $\tau_e$ and $\Delta M$ were determined over more than 10 events and the experimental results are shown in Fig. \ref{fig:experimental_results} and \ref{fig:experimental_results_dmm}.

%Finally, we measured the two characteristic times of a geyser eruption $\tau$ and $\tau_e$ and the exchanged mass $\Delta M$ as a function of $\mathcal{P}$ and $H$

One observes in Fig. \ref{fig:experimental_results}(a) that the elapsed time $\tau$ between two eruptions lasts from 2 to 30 min in our experiments. Also, this characteristic time increases with the geyser height $H$ and decreases with the specific heating power $\mathcal{P}$. The logarithmic scale diagram in the inset of Fig. \ref{fig:experimental_results}(a) reveals that the time $\tau$ between eruptions is inversely proportional to $\mathcal{P}$. Besides, the erupting time $\tau_e$ is between one and two minutes in the present set of experiments reported in Fig \ref{fig:experimental_results}(b). This characteristic time increases with the geyser height $H$ and shows a small dependency with the specific heating power $\mathcal{P}$. One notices in Fig. \ref{fig:experimental_results_dmm} that the exchanged mass fraction $\Delta M / M$ increases with the specific heating power and the water column height.

\section{Model}\label{sec:model}

%This section aim at rationalizing the observations performed previously.

\subsection{Geyser frequency}

The duration between two successive geyser eruptions is approached by estimating the time needed to heat the water in the reservoir from the initial temperature $T_{r,i}$ to the boiling temperature $T_b (H)$ corresponding to a water column of height $H$ [Fig. \ref{fig:t_T_r}(b)]. If a specific heat power $\mathcal{P}$ is transferred to the reservoir, its temperature increases from $T_{r,i}$ to $T_b (H)$ in a time

\begin{equation}
\tau=\frac{c}{\mathcal{P}} \left[ T_b (H) - T_{r,i}\right]
\label{eq:initial_temperature}
\end{equation}

During an eruption, a mass of water $\Delta M$ moves from the reservoir to the pool. The eruption stops when the temperature of the water remaining in the reservoir is close to the boiling temperature $T_b (h)$ corresponding to the pressure imposed by a water column of height $H$ and vaporization weakens. At this moment, the exchanged mass $\Delta M$ moves from the pool to the reservoir and fills it completely. Thus, the temperature $T_{r,i}$ of the reservoir at this particular time is

\begin{equation}
T_{r,i} = \frac{(M - \Delta M) \, T_{b} (h) + \Delta M \, T_p }{M}
\label{eq:mix_temperature}
\end{equation}

Combining Eqs. (\ref{eq:initial_temperature}) and (\ref{eq:mix_temperature}) yields

\begin{equation}
\tau=\frac{c}{\mathcal{P}} \left[ \frac{\Delta M}{M} (T_b (h) - T_p) + (T_b (H) - T_b (h)  )\right]
\label{eq:geyser_period}
\end{equation}

One observes that Eq. (\ref{eq:geyser_period}) gives two contributions for the characteristic time between geyser eruptions. The first contribution corresponds to the cool down of the reservoir by the entry of a mass $\Delta M$ of water coming from the pool maintained at a temperature $T_p$. The second contribution is the time required to reach the boiling temperature $T_b (H)$ below a water column of height $H$. These contributions imply that the geyser height $H$, the pool temperature $T_p$ and the geometry of the plumbing system impact the time between eruptions. In our experiments, all quantities involved in Eq. (\ref{eq:geyser_period}) are known except $\Delta M / M$ which has been measured, thus an estimation of the time between eruptions $\tau$ can be deduced. Such predictions are plotted by the way of solid lines in Fig. \ref{fig:experimental_results}(a) for the two geyser heights inspected experimentally ($H=1.60 \, \rm{m}$ and $H=11.50 \, \rm{m}$). The good agreement between experiments and predictions validate the approach considered above.

\subsection{Erupting time}\label{sec:erupting_time}

After the beginning of an eruption, the temperature of the reservoir is $T_b (H)$ whereas the overlying pressure is close to the atmospheric one and imposes an equilibrium temperature $T_b (h)$. The erupting time is associated with the time needed to cool down the reservoir from $T_b (H)$ to $T_b (h)$ by vaporizing water [stage (c) in Fig. \ref{fig:steps}]. In our situation, we assume that this process is limited by the thermal transfer occurring in the reservoir with bubbles formation on nucleation sites, their growth and their ascension. Such a thermal transfer depends on the difference between the temperature of the water in the reservoir and the boiling temperature corresponding to the imposed pressure. To our knowledge, such a phenomenon has never been described quantitatively. Thus, we assume empirically a power law dependency for the power rate lost by the superheated water, \textit{i.e.} $\dot{Q}= \alpha (T_r - T_b (h))^\beta$. In first approximation, the power lost by the vaporization is balanced by the temperature variation of the reservoir $M c \, \dot{T}_r = - \alpha (T_r - T_b (h))^\beta$. Solving the previous equation from a reservoir temperature $T_b (H)$ to a threshold temperature $T_{r , t}$, for which an eruption stops, provides the following prediction for the time of eruptions

%\begin{equation}
%\tau_e = \frac{1}{\beta} \ln \left( \frac{T_b (H) - T_b (h) }{T_{r,t} - T_b(h)} \right)
%\label{eq:geyser_tau_e}
%\end{equation}

\begin{equation}
\tau_e = \frac{\alpha}{Mc (1-\beta)} \left[ (T_b (H) - T_b (h))^{1- \beta} -  (T_{r,t} - T_b (h))^{1- \beta} \right]
\label{eq:geyser_tau_e}
\end{equation}

One notices that if $\beta<1$, the erupting time predicted by Eq. (\ref{eq:geyser_tau_e}) increases with the geyser height as observed experimentally in Fig. \ref{fig:experimental_results}(b).  More quantitatively, the previous model can be adjusted on experimental data. Solid lines in Fig. \ref{fig:experimental_results}(b) correspond to the prediction of Eq. (\ref{eq:geyser_tau_e}) for $\beta= 0.66$, $\alpha = 38$ SI and $T_{r,t} - T_b (h)=0.1 \, \rm{^\circ C}$. The deviation of experimental data from the theoretical predictions at large heating power ($\mathcal{P}>150$ W/kg) can be attributed to the omission of the energy injected by the heating plate in the energy balance of the model. Indeed, the specific heating power $\mathcal{P}$ delays the cooling of the water in the reservoir by vaporization.\\

\subsection{Exchanged mass}\label{sec:exch_mass}

\begin{figure}[h!]
\centering
\includegraphics[width=3.5cm]{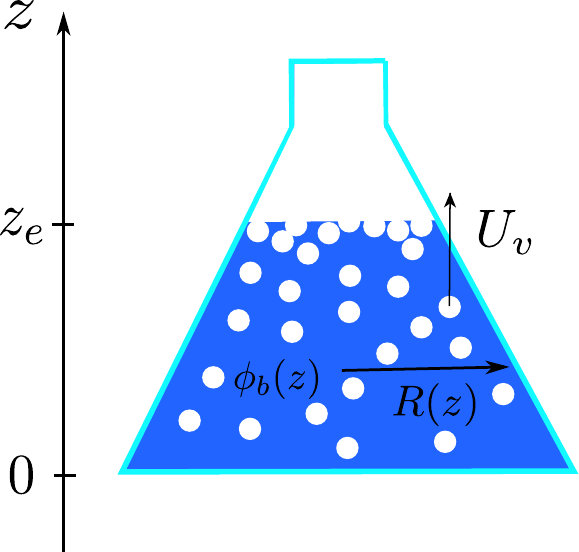}
\caption{Definition of the parameters used in the model developed in section \ref{sec:exch_mass}. All the vapor bubbles have the same upwards velocity $U_v$. Due to the reduction of the reservoir radius $R(z)$ with increasing height $z$, the bubble volume fraction $\phi_b (z)$ increases along this direction up to a critical volume fraction $\phi_{b,c}$ which defines the equilibrium position $z_e$ of the liquid/gas interface.}
		\label{fig:sketch_bubbles}
\end{figure}

When an eruption has occurred and water has left the vent, the vaporization in the reservoir results from both the injected heat power $\mathcal{P}M$ and the heat flux $\dot{Q}$ coming from the water superheat. Thus, the power balance in such a situation provides: $\mathcal{L} \dot{M}= \mathcal{P} M + \alpha (T_b (H) - T_b (h))^\beta $ where $\dot{M}$ is the mass flow rate of vapor produced in the reservoir and $\mathcal{L}$ the latent heat of water. The vaporization generates bubbles which grow and rise in the reservoir. Due to the conical shape of the reservoir, the bubble volume fraction $\phi_b (z)$ increases along the vertical axis $z$ as sketched in Fig. \ref{fig:sketch_bubbles}. When the volume fraction reaches a critical value $\phi_{b,c}$, the bubbles coalesce and finally form the vapor phase. Considering that the critical bubble volume fraction corresponds to the fraction for which a single bubble cannot be added to a cubic close-packed arrangement of monodispered bubbles, we estimate that $\phi_{b,c}=0.33$. Thereby, the equilibrium position $z_e$ of the liquid/vapor interface during an eruption can be deduced from the critical bubble volume fraction in the liquid. Introducing the bubbles upwards velocity $U_b$ in the reservoir, the vapor density $\rho_v$ and $R(z)$ is the radius of the reservoir at the altitude $z$ relatively to the bottom of the reservoir, we get: $\mathcal{L} \, \rho_v \, \phi_b (z)  \, \pi R^2 (z) \, U_b = \mathcal{P} M + {\alpha (T_b(H) - T_b (h))^\beta}$. Haberman and Morton showed that the terminal velocity of vapor bubbles in water is between 20 cm/s and 30 cm/s for bubble of radius comprised between 0.5 mm and 15 mm \cite{haberman1953experimental}. For the sake of simplicity, we consider that all the bubbles in the vent rise at the same velocity $U_b \simeq 25$ cm/s, a fact which is consistent with our observations. The critical bubble volume fraction $\phi_{b,c}$ is reached for a reservoir radius 

%When the reservoir reaches its boiling point, the injected heat vaporizes water and forms bubbles. As the bubbles grow and rise, the cross section of the reservoir reduces and thus the bubble volume fraction increases with the height. In the end, the bubble volume fraction reaches a critical value above which the overlying water cannot flow down and is entrained by bubbles motion. The critical height determines the stable position of the liquid/vapor interface during eruptions and the mass exchanged between the reservoir and the pool during this phase. Considering that the heat lost $Q$ plus the injected heat $\mathcal{P}M$ vaporize water, we get: $L \dot{M}= \alpha (T_b (H) - T_b (h))^\beta + \mathcal{P} M$ where $\dot{M}$ is the mass flow rate of vapor produced in the reservoir and $L$ the specific heat of vaporization of water. Introducing the bubble volume fraction $\phi_b(z)$ and the reservoir radius $R(z)$ at the altitude $z$ from the bottom of the water column and assuming a constant bubbles upwards velocity $U_b$, the previous equality provides: $\phi_b (z) \pi R^2 (z) U_b = ( {\alpha (T_b(H) - T_b (h))^\beta} + \mathcal{P} M ) /{L \rho_v}$. The equilibrium position $z_e$ of the liquid/vapor interface during an eruption is determined by the critical bubble volume fraction $\phi_{b,c}$ above which the bubbles clog in the vent and entrain the upper part of water. Thus, $z_e$ verifies the relation

\begin{equation}
R(z_e) = \sqrt{ \frac{\mathcal{P} M + \alpha (T_b(H) - T_b (h))^\beta  }{ \pi \, L \, \rho_v \, \phi_{b,c} \, U_b }}
\label{eq:mass_exchanged}
\end{equation}

We deduce from Eq. (\ref{eq:mass_exchanged}) and geometrical considerations, the mass of water $\Delta M$ ejected out of the reservoir during an eruption for the parameters corresponding to the experiments of section \ref{sec:experiments} ($M=1.2$ kg, $\rho_v=0.9 \, \rm{kg/m^3}$, $\mathcal{L}=2.6 \times 10^6$ J/kg, $\beta=0.66$ and $\alpha=38$ SI) and a critical bubble volume fraction $\phi_{b,c}=0.33$. The prediction of the theory is show in Fig. \ref{fig:experimental_results_dmm} by the way of solid lines. One observes the relative agreement between the theory and experimental data.

% Considering the value of $\alpha$ determined in section \ref{sec:erupting_time} ($\alpha=206$ W/K), $T_b(H) - T_b (h)=2 \, \rm{^\circ C}$ , $L=2.3 \times 10^6 \, \rm{J/kg}$, $\rho_v=0.9 \, \rm{kg/m^3}$, $U_b = 0.25 \, \rm{m/s}$ and $\phi_{b,c} \simeq 0.3$, we get $S(z_e)=22 \, \rm{cm^2}$. This cross section corresponds to a radius $R(z_e)=2.6$ cm. 

\section{Conditions for geysers existence}\label{sec:discussion}

The aim of this section is to inspect under which conditions a toy geyser can lead to periodic eruptions. First, we inspect the physical origin of toy geyser eruptions, and second we consider the domain of existence of three non-periodic regimes: the fumarole regime, the hot springs regime and the boiling pool regime. 

\subsection{Geyser's birth: bubble clogging}

First, we investigate the effect of the vent radius on the toy geyser. Experimentally, we observed that if the vent radius is too large ($R>20$ mm), the vapor bubbles produced by the boiling in the reservoir are able to rise without entraining the surrounding water and no eruptions occur. Thus, the birth of a geyser eruption is caused by the transition from a dispersed flow regime to an annular flow regime in the vent. The regimes of a two-phase flow in a vertical pipe have been studied extensively by Mishima \textit{et al.} \cite{mishima1996some}. The transition between a dispersed and an annular flow in the vent can be approached with a critical air volumic fraction above which bubbles clog and entrain the surrounding water. In the present situation, the vapor volumic fraction in the vent $\phi_b$ can be estimated by assuming that when the reservoir reaches its boiling temperature $T_b (H)$,  all the specific heating power $\mathcal{P}$ is used to vaporize water. The balance between the vapor departure through the vent and the vapor production in the reservoir yields

\begin{equation}
\phi_b U_b \pi R^2 = \frac{\mathcal{P} M }{\mathcal{L} \, \rho_v}
\label{eq:bubble_flow}
\end{equation}

Using the critical vapor volume fraction $\phi_{b ,c}$ introduced in section \ref{sec:exch_mass}, we predict the critical vent radius $R_c$ below which eruptions occur

\begin{equation}
R_c = \sqrt{\frac{\mathcal{P} M}{\pi \mathcal{L} \rho_v U_b \phi_{b,c}}}
\label{eq:radius_clogging}
\end{equation}

For $\mathcal{P}=220 \, \rm{W/kg}$, $M=1.2 \, \rm{kg}$, $\mathcal{L}=2.3 \times 10^6 \, \rm{J/kg}$, $\rho_v=0.9 \, \rm{kg/m^3}$, $U_b = 0.25 \, \rm{m/s}$ and $ \phi_{b,c}=0.33$, we get $R_c=17 \, \rm{mm}$. The vent radius used in the experiments presented in section \ref{sec:experiments} ($R= 3.0 \, \rm{mm}$) is smaller than $R_c$ which explains why periodic eruptions were observed. In order to verify the previous criteria, we built a model geyser with the same reservoir and the same height but a different vent radius $R=20 \, \rm{mm}$ and no eruptions have been observed, consistently with our approach. Such a regime is called the \og hot spring \fg  regime because only tiny and hot vapor bubbles reach the pool.

Beyond the case of a toy geyser, bubble clogging in a vertical constriction is a phenomenon observed in a large variety of situations such as volcanic eruptions \cite{vergniolle2000hawaiian}, cold CO$\,_2$ geysers \cite{han2013characteristics}, mentos in diet coke \cite{coffey2008diet} and beer taping \cite{rodriguez2014physics}. In the particular case of volcanic eruptions, the elevation of the internal pressure in the volcano and thus the explosiveness of the eruption depend on possibility for gas to escape. Depending on the volatile content of the magma, its viscosity and the configuration of the magmatic chamber, different types of eruption occur from Hawaiian to Plinian eruptions \cite{walker1973explosive,wilson1980relationships}.

%\textcolor{red}{Can we try to discuss the effect of the surface tension or the pressure on the vapor bubble size and predict how it will impact the previous transition?}

\subsection{Geyser's death: transition to fumarole}\label{sec:fumarole}

For large specific heating power $\mathcal{P}$, we observed a first eruption followed by a continuous vaporization of the water in the reservoir. In such a case, the liquid in the pool is unable to go down into the reservoir. Because of permanent emission of vapor through the vent, this situation is called the \og fumarole\fg regime. Besides, we have noticed that a toy geyser working in a fumarole regime can be brought back to a periodic regime by increasing the height $h$ of the water in the pool. The minimal pool height which leads to a periodic regime can be discussed quantitatively. During an eruption, when the water in the reservoir has reached its equilibrium temperature $T_b (h)$, the specific heating power $\mathcal{P}$ only induces water vaporization. Thus, the mass of water vapor produced per unit time is $\dot{M}= \mathcal{P}M/\mathcal{L}$ and imposes a mean gas velocity in the vent $U_v= \mathcal{P} M/ \pi R^2 \mathcal{L} \rho_v$. If the dynamic pressure $\rho_v U_v ^2/2$ associated to the vapor flow at the exit of the vent is larger than the hydrostatic pressure $\rho_l g h$ (where $\rho_l$ is the water density) at the bottom of the pool, the water will not re-entry in the vent and the system will stay in a fumarole regime. The critical specific heating power $\mathcal{P}_c$ above which the system does not follows periodic eruptions is

\begin{equation}
\mathcal{P}_c = \pi \mathcal{L} R^2 \sqrt{\rho_v \rho_l  \, g \, h}/ M
\label{eq:power_threshold_fumarole}
\end{equation}

For $R=3.0 \, \rm{mm}$, $h=40 \, \rm{mm}$, $\mathcal{L}=2.3 \times 10^6 \, \rm{J/kg}$, $\rho_v=0.9 \, \rm{kg/m^3}$, $\rho_l=960 \, \rm{kg/m^3}$ and $M = 1.2 \, \rm{kg}$, this relation gives a critical value of $\mathcal{P}_c \simeq 1000 \, \rm{W/kg}$. In the experiments presented in section \ref{sec:experiments}, the maximal specific heat power is $\mathcal{P}=220 \, \rm{W/kg}$ which explains why a geyser regime was always observed. In order to test the validity of Eq. (\ref{eq:power_threshold_fumarole}), we decrease the height $h$ of the water in the pool till the transition towards a fumarole regime occur. For $\mathcal{P}=220 \, \rm{W/kg}$, we found a critical height of water $h=30$ mm which is consistent with the previous law. 

%\textcolor{red}{What can be corrected in this model? I think the criteria at the top of the vent is not the good one but what else can we try?}

\subsection{Geyser's death: convection}

In the description of the toy geyser done previously, we have considered that the heat power injected in the system is entirely transferred to the reservoir. Actually, thermal convection can transfer a part of the injected heat power from the reservoir to the vent and to the pool. The thermal energy per unit time conveyed in the vent can be expressed as $k (T_r - T_b) \pi R^2$ with $k$ the heat transfert coefficient in the case of vertical natural convection in a vertical pipe heated from below. This coefficient depends on the Rayleigh number and the Grashof number but stays between 100 and 1000 $\rm{W.K^{-1}.m^{-2}}$ in the range of parameter encountered in our experiments \cite{jackson1989studies}. The limit case where the convective flux is equal to the injected heat power $\mathcal{P} M$ occurs for a critical vent radius

\begin{equation}
R_c ' = \sqrt{\frac{\mathcal{P} M}{\pi k (T_r - T_p)}}
\label{eq:radius_convection}
\end{equation}

For $\mathcal{P}=220 \, \rm{W/kg}$, $M=1.2 \, \rm{kg}$, $k= 500 \, \rm{W.K^{-1}.m^{-2}}$ and $T_r - T_p= 70 \, \rm{^\circ C}$ we get $R_c '=41 \, \rm{mm}$. This approach explains why thermal convection in the vent has been neglected in the experiments presented in section \ref{sec:experiments} for which $R=3.0 \, \rm{mm}$. Indeed, in such a situation the convective flux is about 200 times smaller than the injected heat power. Finally, one expects that for a vent radius larger than $R_c '$, the injected heat power will be transferred to the entire volume of water and boiling will occur everywhere. This regime of the toy geyser is called the \og boiling spring\fg regime in the following of this paper.

\subsection{Phase diagram}\label{sec:phase_diagram}

\begin{figure}[h!]
\centering
\includegraphics[width=12cm]{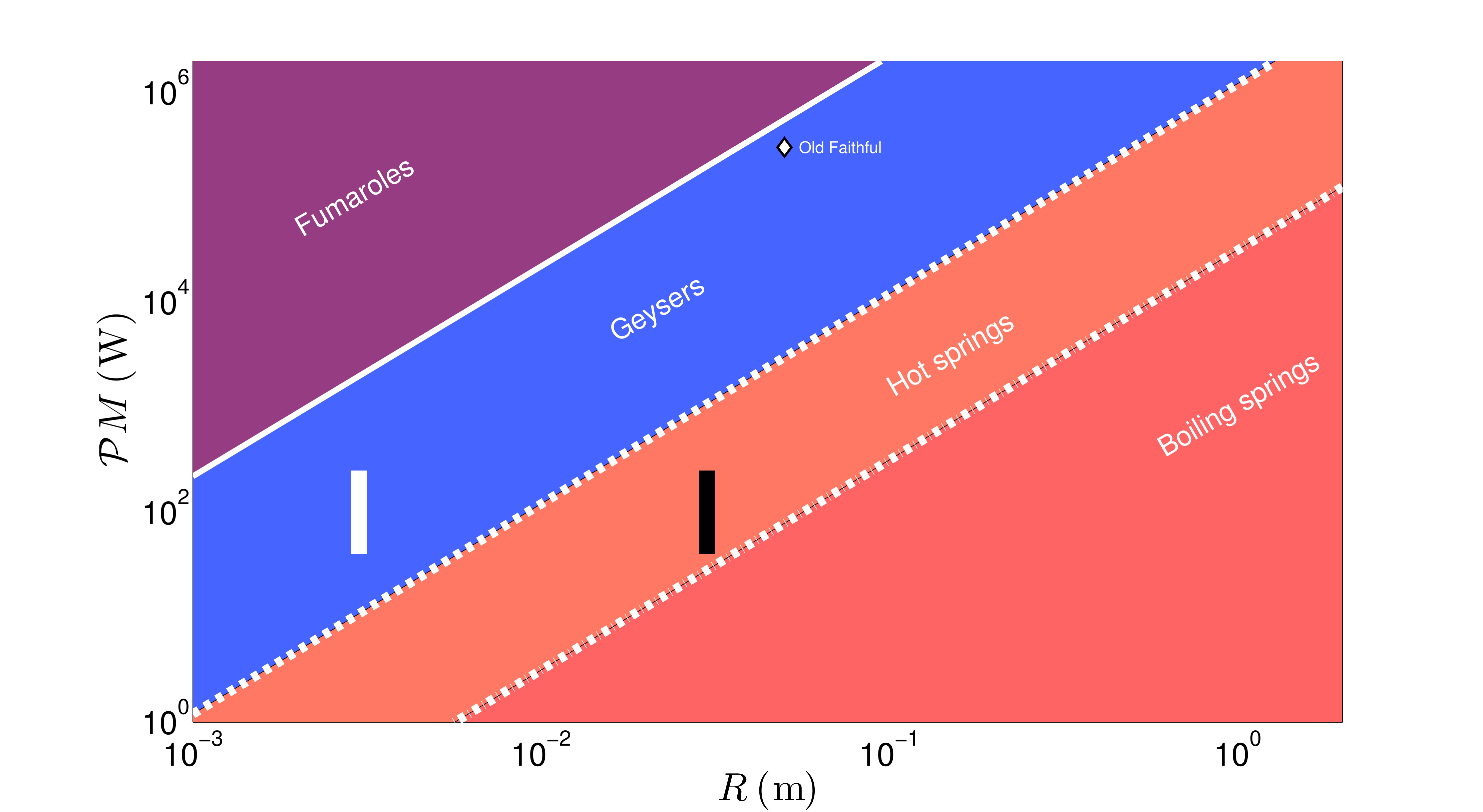}
\caption{Phase diagram of the behavior of a toy geyser as sketched in Fig. \ref{fig:geyser}(c) as a function of the vent radius $R$ and the injected heat power $\mathcal{P} M$. The white solid line represents transition between a fumarole and a geyser regime as predicted by Eq. (\ref{eq:power_threshold_fumarole}). The dashed line indicates the limit between the geyser and the hot spring regimes according Eq. (\ref{eq:radius_clogging}). The dotted line corresponds to the transition between a hot spring and a boiling spring regime as predicted by Eq. (\ref{eq:radius_convection}). The white vertical stripe indicates the range of parameter inspected experimentally in Section \ref{sec:experiments} where a geyser regime has been reported. On the contrary, the dark stripe shows the range of parameter inspected experimentally in section \ref{sec:fumarole} and for which a hot spring regime has been observed. Also, the estimations of the mean heating power and minimal radius of the constriction for the Old Faithful are reported in the diagram by the way of a diamond.}
\label{fig:Phase_diagramme}
\end{figure}

The three transitions discussed previously between the different regimes of a toy geyser can be gathered on a single diagram which is presented in Fig. \ref{fig:Phase_diagramme}. Such a diagram shows the behavior of a toy geyser depending on the injected heat power $\mathcal{P} M$ and the vent radius $R$. If the specific heating power is larger than the critical one $\mathcal{P}_c$ corresponding to Eq.  (\ref{eq:power_threshold_fumarole}), the set-up exhibits a fumarole regime. Differently, if the vent radius is between the two critical radii $R_c$ and $R_c '$ predicted by Eq. (\ref{eq:radius_clogging}) and (\ref{eq:radius_convection}), the system behaves as a hot spring. Finally, if $R>R_c'$ than the water in the reservoir convects and the toy geyser adopts a boiling spring regime. In between these regimes, lies the regime corresponding to geyser periodic eruptions which required $R<R_c$ and $\mathcal{P}<\mathcal{P}_c$. Also, we report in Fig. \ref{fig:Phase_diagramme} the range of experimental parameters used for the experiments detailed in Section \ref{sec:experiments} and Section \ref{sec:fumarole} where a geyser regime and a hot spring regime have been reported. 

As underlined in the introduction, most of natural geysers adopt a complex dynamics rather than a precise frequency of eruption. Such behavior may be attributed to a change in external parameters such as the water inflow, the atmospheric temperature and pressure, tidal forces and tectonic stress \cite{rinehart1972fluctuations} or be the result of the internal mechanism of geysers. Indeed, the interconnection between the plumbing systems of different geysers is known to induce a chaotic and complex dynamics of eruptions \cite{nicholl1994old}. Despite the numerous differences that distinguish toy geysers from natural ones, it is interesting to observe where natural geysers are located in the previous phase diagram. For this purpose, we collected in the literature estimations of the minimal radius of constriction and the mean injected heat power of the Old Faithful in Yellowstone \cite{vandemeulebrouck2013plumbing,rinehart1980geysers} and report them in Fig. \ref{fig:Phase_diagramme}. One notices that the parameters corresponding to this natural geyser is in the domain where a toy geyser adopts a periodic regime. We hope that future characterizations of geothermal systems will allow to compare more rigorously this theoretical phase diagram with experimental facts. Such a procedure can be a way to determine if natural geysers work on the same principle than toy geysers.

\section{A step toward the complexity of natural geysers}\label{sec:two_reservoirs}

\begin{figure}[h!]
\centering
\includegraphics[width=6cm]{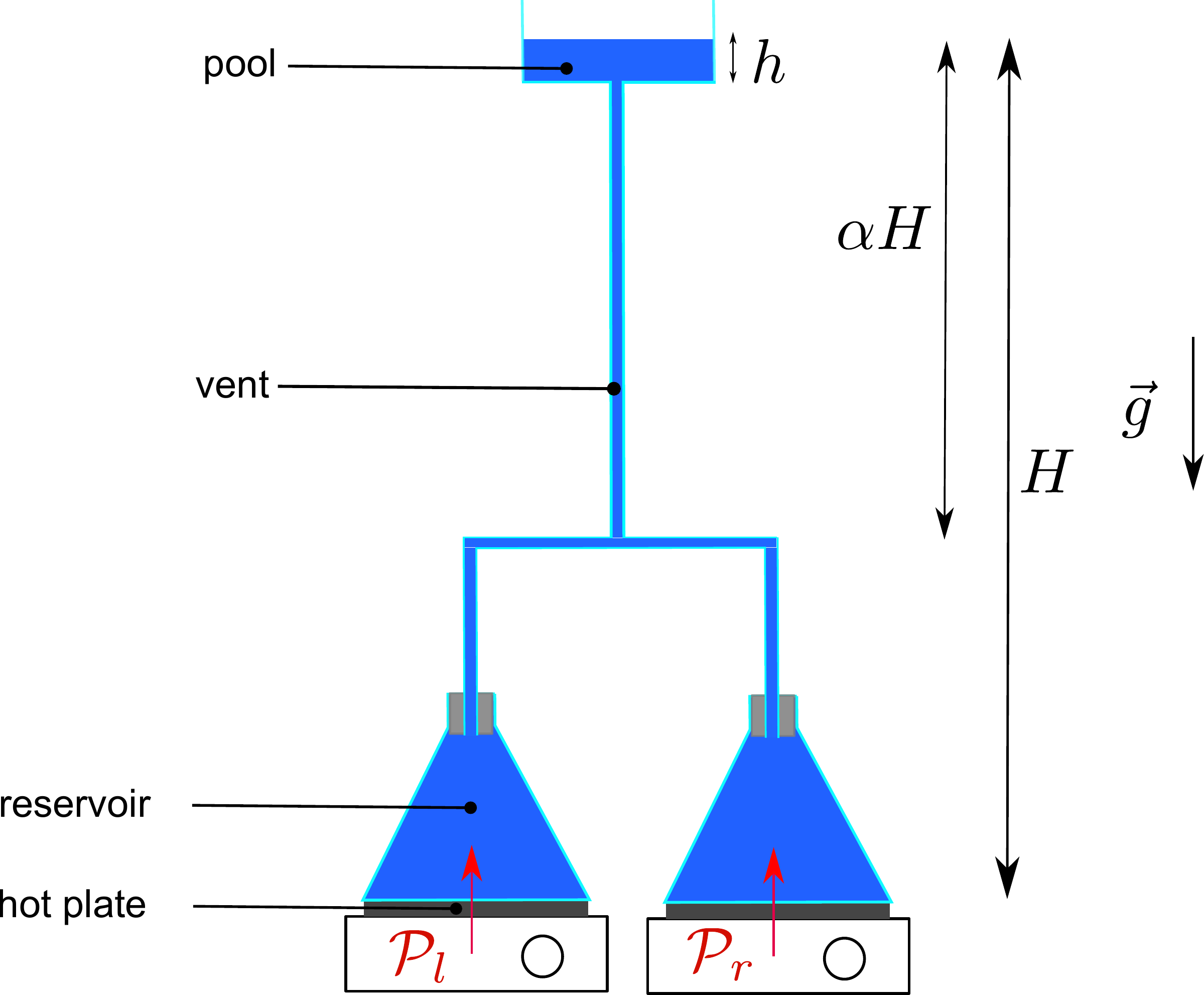}
\caption{Sketch of a toy geyser experiment with two interconnected reservoirs. The specific heating power received by each reservoir is respectively $\mathcal{P}_l$ and $\mathcal{P}_r$. The two vents are connected at a height $\alpha H$ proportional to the total height $H$ of the system.}
\label{fig:setup_2}
\end{figure}

A way to approach the complexity of natural geysers is to build a set-up with two interconnected reservoirs. Both reservoirs received a different specific heating power and they are connected to a single pool as sketched in Fig. \ref{fig:setup_2}. The connection between the vents of the two reservoirs is made at a fraction $\alpha$ of the total height $H$ of the water column.

\subsection{Experimental data}\label{sec:double_exp}

\begin{figure}[h!]
\centering
	\begin{minipage}[c]{0.5\textwidth}
  		\centering
  		\hspace{0.4cm}(a)\\
		\vspace{0.1cm}
		\includegraphics[width=7cm]{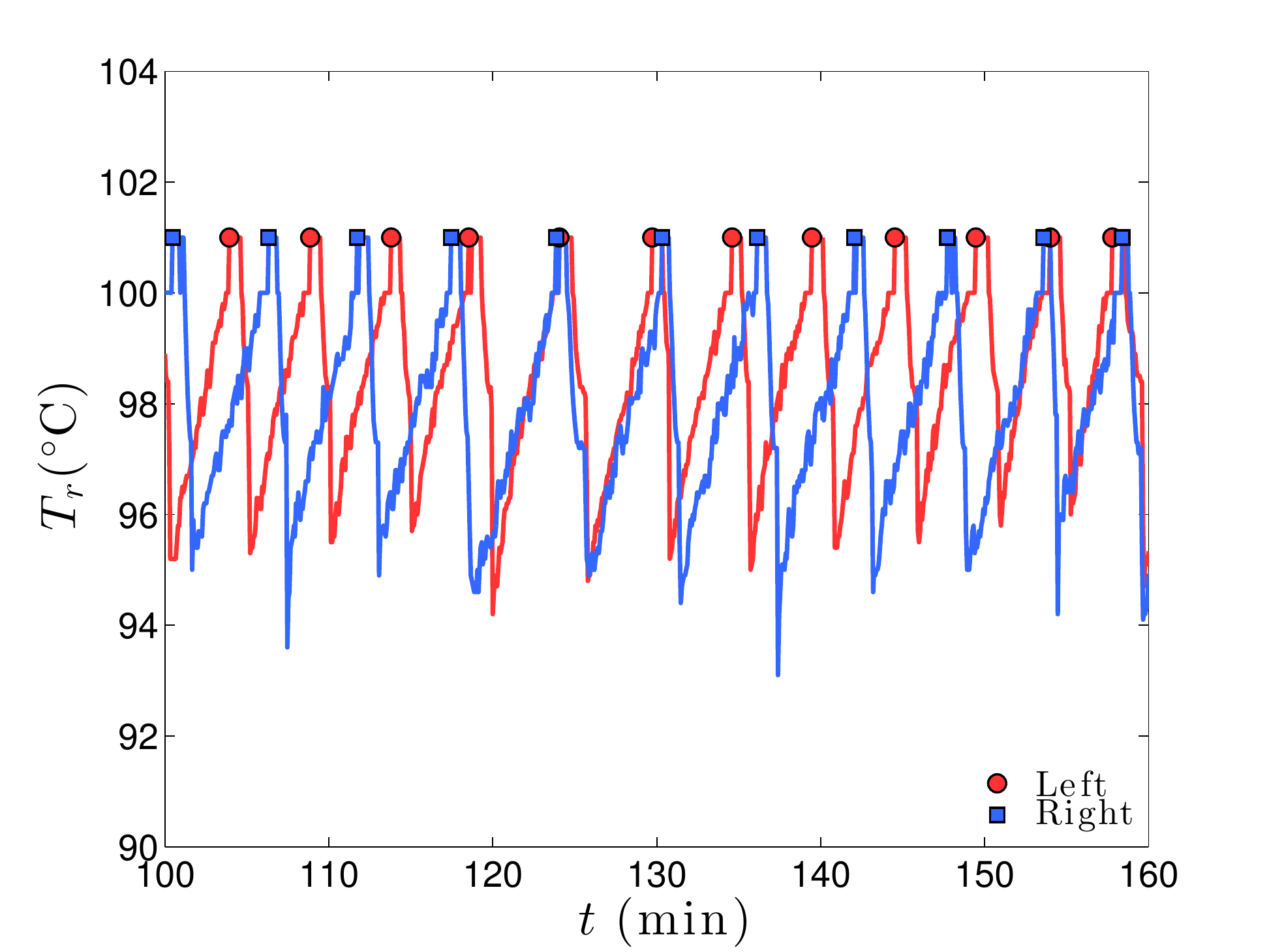}
 	\end{minipage}%
	\begin{minipage}[c]{0.5\textwidth}
  		\centering
  		\hspace{0.4cm}(b)\\
		\vspace{0.1cm}
		\includegraphics[width=7cm]{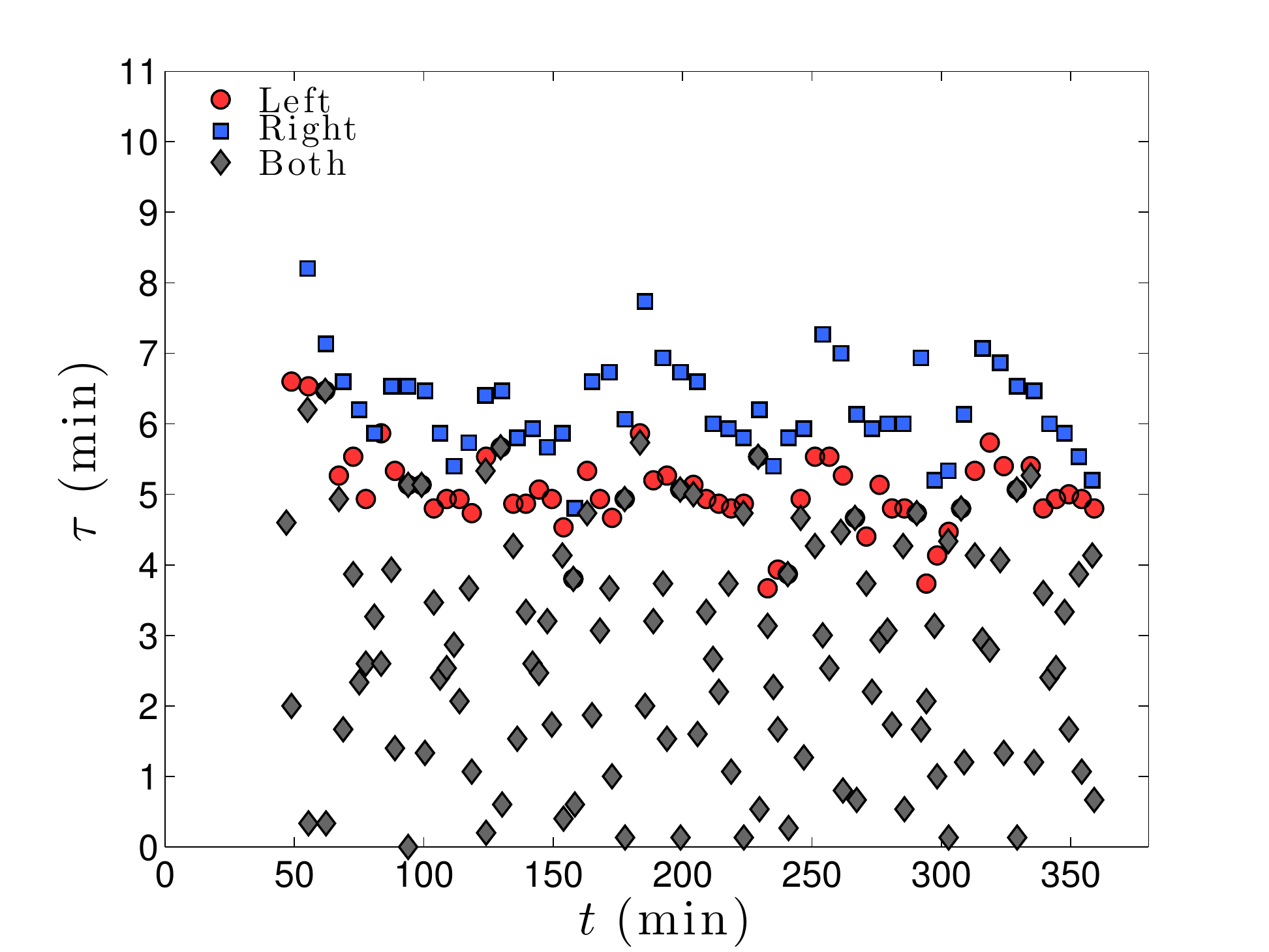}
 	\end{minipage}%
 	\caption{Evolution over time of an interconnected toy geyser as sketched in Fig. \ref{fig:setup_2}. The total height of the geyser is $H=1.60$ m and $\alpha=0.20$. The specific heat power injected in the left and right reservoirs are respectively $\mathcal{P}_l = 110$ W/kg and $\mathcal{P}_r = 100$ W/kg. (a) Measured temperatures as a function of time in each reservoir. Symbols indicate the time of eruptions. (b) Times $\tau$ between eruptions over time for each reservoir (red dots and blue square) and for the global system (gray diamonds).}
		\label{fig:double_exp}
\end{figure}

Figure \ref{fig:double_exp}(a) shows the evolution over time of the temperature in each reservoir of an interconnected toy geyser. One notices that the saw-tooth evolution of the temperature in a reservoir is not periodic anymore differently from the case of a geyser with a single reservoir inspected in section \ref{sec:experiments}. The lost of regularity is underlined in Fig. \ref{fig:double_exp}(b) where the times between eruptions for each reservoir and for the total system are presented as a function of time. The aim of this section is to describe and rationalize the dynamics of toy geyser with two coupled reservoirs.  

\begin{figure}[h!]
\centering
	\begin{minipage}[c]{0.3\textwidth}
  		\centering
  		\hspace{0.4cm}(a)\\
		\vspace{0.05cm}
		\includegraphics[width=5cm]{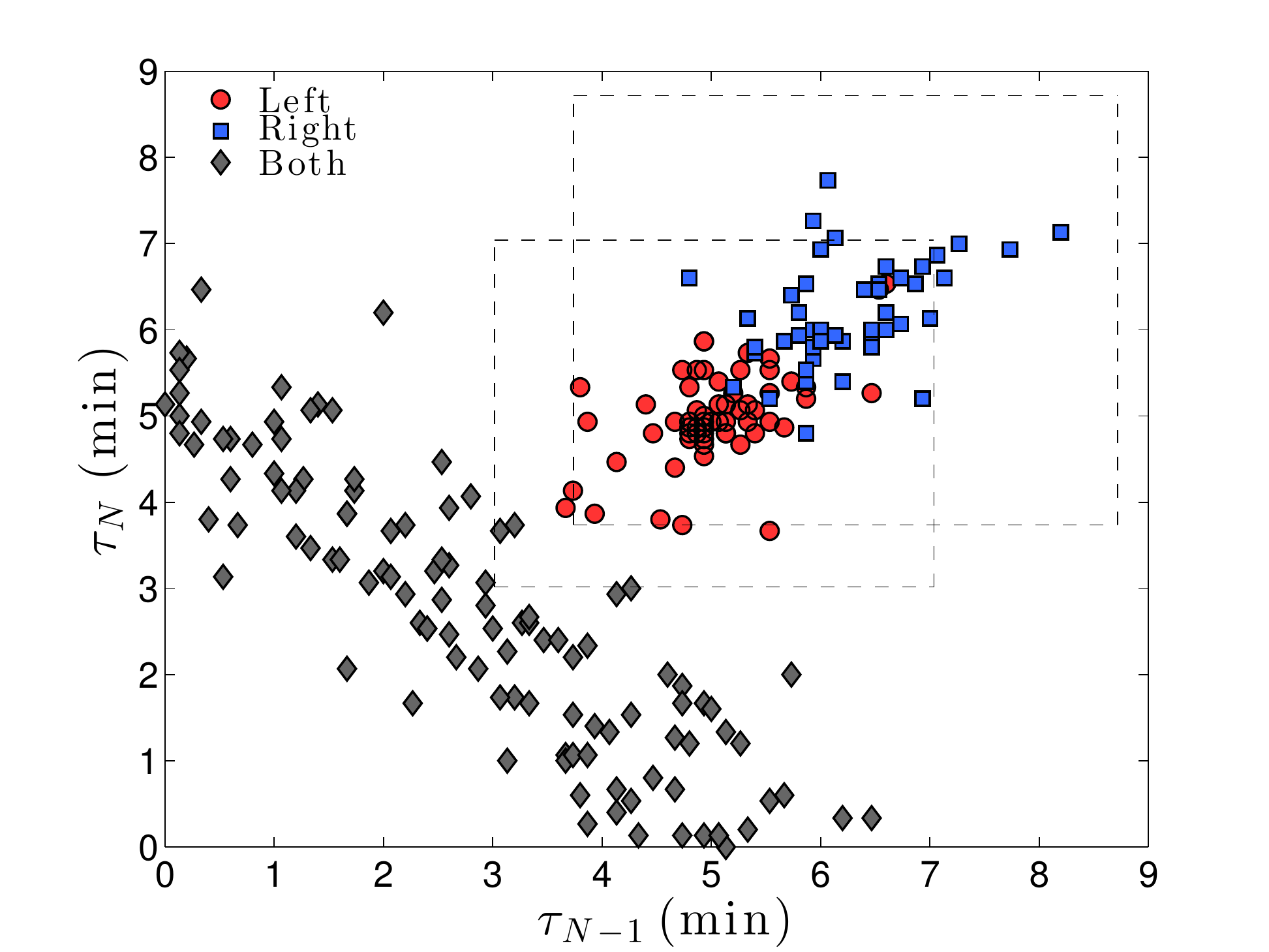}
 	\end{minipage}%
	\begin{minipage}[c]{0.3\textwidth}
  		\centering
  		\hspace{0.4cm}(b)\\
		\vspace{0.05cm}
		\includegraphics[width=5cm]{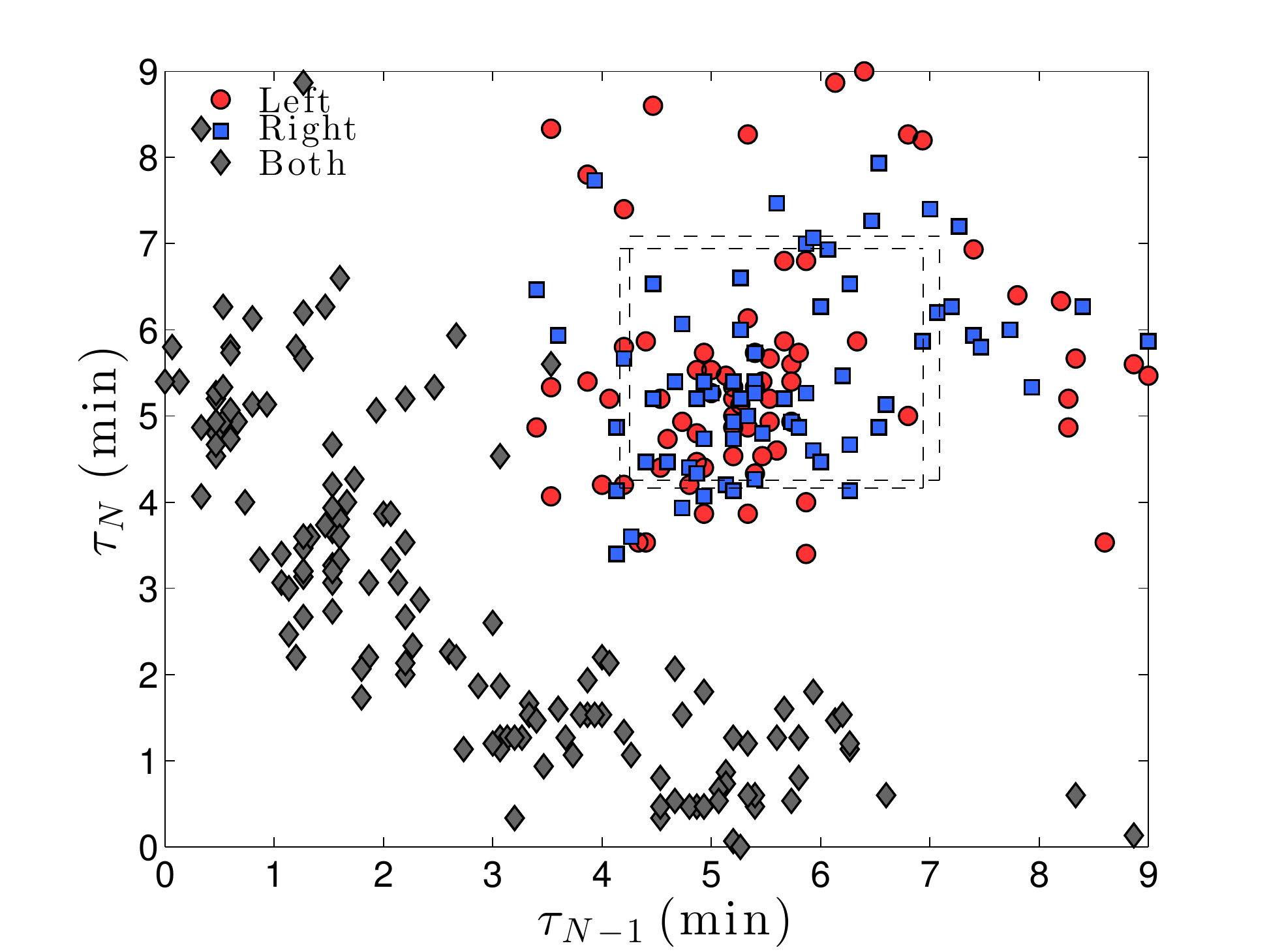}
 	\end{minipage}%
	\begin{minipage}[c]{0.3\textwidth}
  		\centering
  		\hspace{0.4cm}(c)\\
		\vspace{0.05cm}
		\includegraphics[width=5cm]{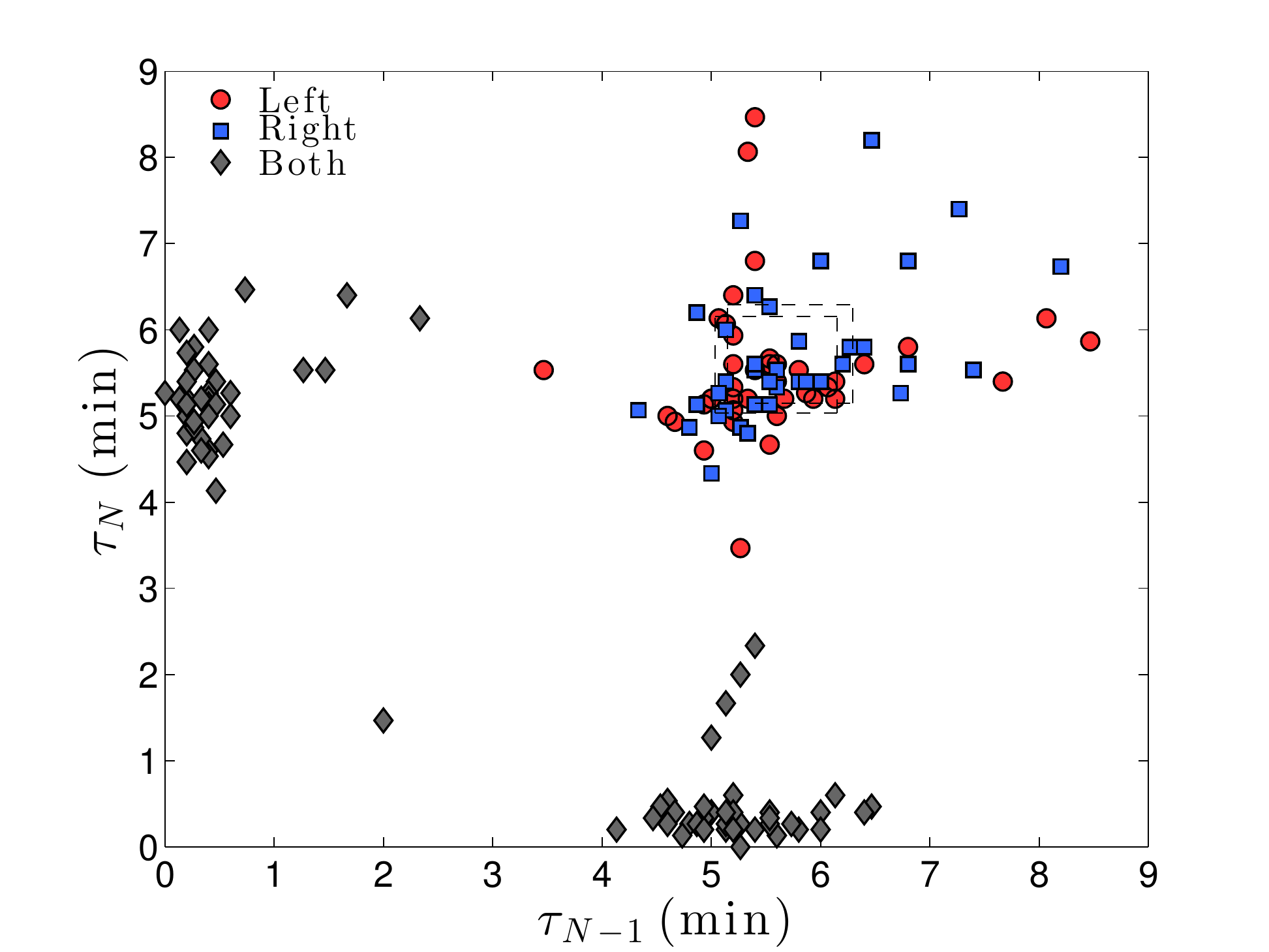}
 	\end{minipage}%
 	\caption{Correlation of time between eruptions of the event numbered $N$ and the previous event numbered $N-1$. Red dots and blue squares indicate the time between eruption in each reservoir whereas gray dots represent the time between eruptions for the total system. (a) $\alpha=0.20$. (b) $\alpha=0.70$. (b) $\alpha=0.90$. The dashed rectangles show the predictions for the dispersion for the time between eruptions based on the model introduced in Section \ref{sec:model_double}.}
		\label{fig:double_exp2}
\end{figure}

A different way to present the data of Fig. \ref{fig:double_exp}(b) is to plot the time between eruption of an event numbered $N$ as a function of the previous time between eruption (numbered $N-1$). Such a procedure allows to visualize the correlation between successive eruptions and is done in Fig. \ref{fig:double_exp2} for three different coupling parameters $\alpha$. The figure shows the correlation for each reservoir (red dots and blue squares for left and right reservoirs respectively) and for the total system (gray diamonds). One observes in Fig. \ref{fig:double_exp2}(a) that even for small coupling parameter $\alpha=0.20$, the time between eruptions in each reservoir undergoes a large dispersion (about two minutes of dispersion while the mean period is about six minutes). Besides, the data for the whole system (gray diamonds) occupy a trapezoidal shaped domain. When the coupling parameter increases to $\alpha=0.70$ [Fig. \ref{fig:double_exp2}(b)], the dispersion for each reservoir increases and the domain of the global eruption adopt a \og banana \fg shape. Finally, for large coupling parameter $\alpha=0.90$, Fig. \ref{fig:double_exp}(c) indicates a reduction of the dispersion of the time between eruptions in each reservoir and for the whole system.

\subsection{Model}\label{sec:model_double}

\begin{figure}[h!]
\centering
	\begin{minipage}[c]{0.3\textwidth}
  		\centering
  		\hspace{0.4cm}(a)\\
		\vspace{0.05cm}
		\includegraphics[width=5cm]{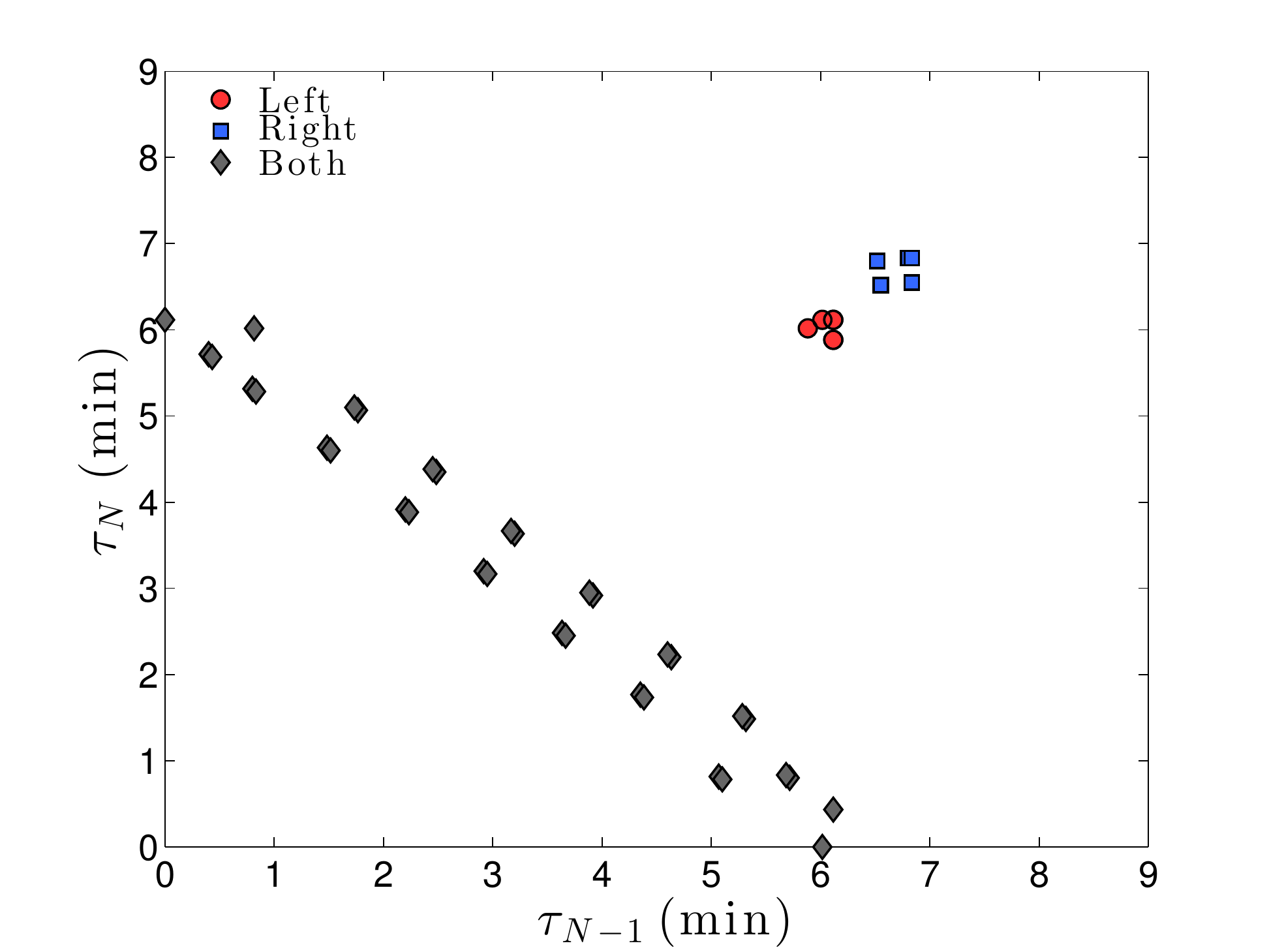}
 	\end{minipage}%
	\begin{minipage}[c]{0.3\textwidth}
  		\centering
  		\hspace{0.4cm}(b)\\
		\vspace{0.05cm}
		\includegraphics[width=5cm]{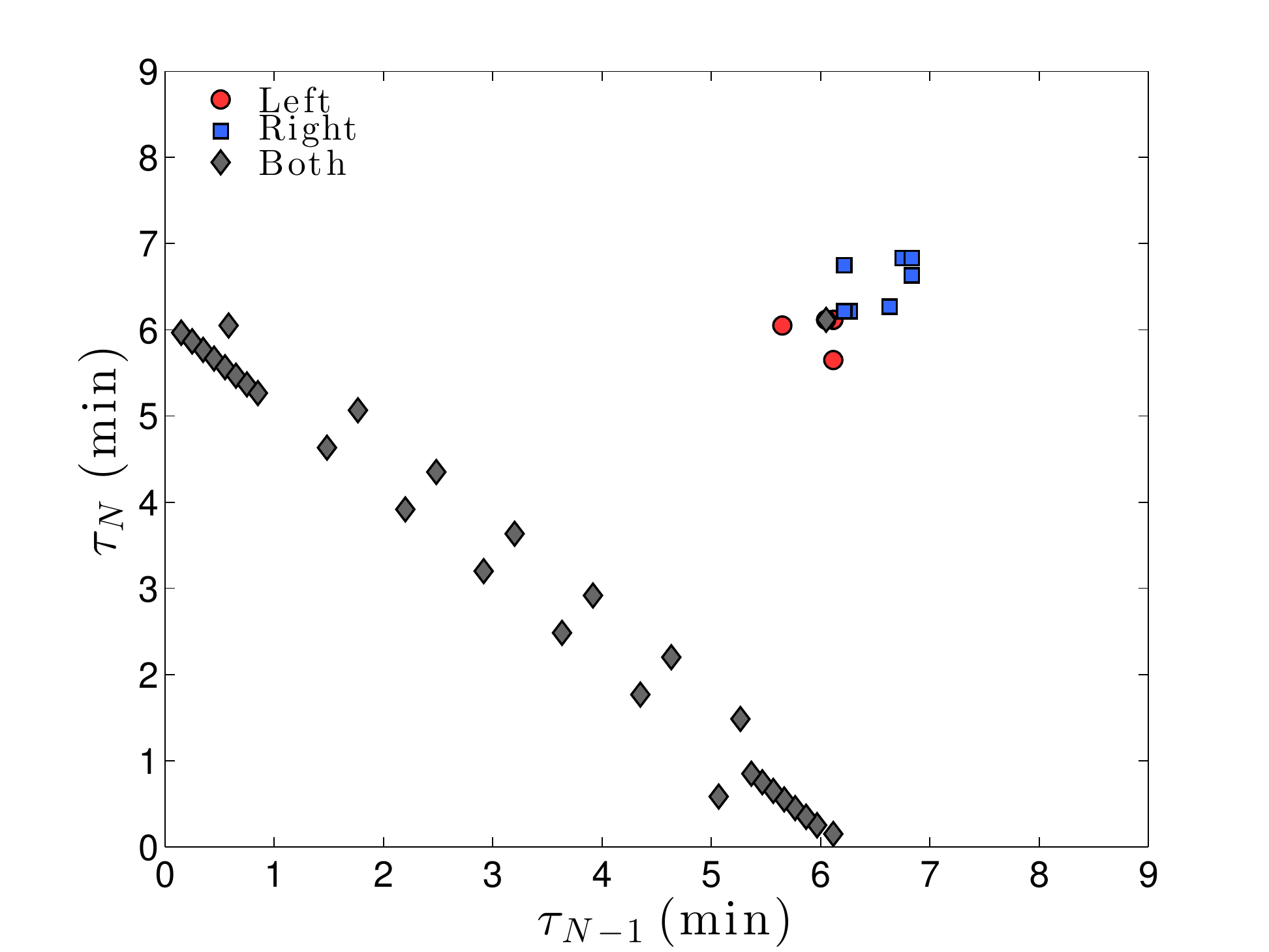}
 	\end{minipage}%
	\begin{minipage}[c]{0.3\textwidth}
  		\centering
  		\hspace{0.4cm}(c)\\
		\vspace{0.05cm}
		\includegraphics[width=5cm]{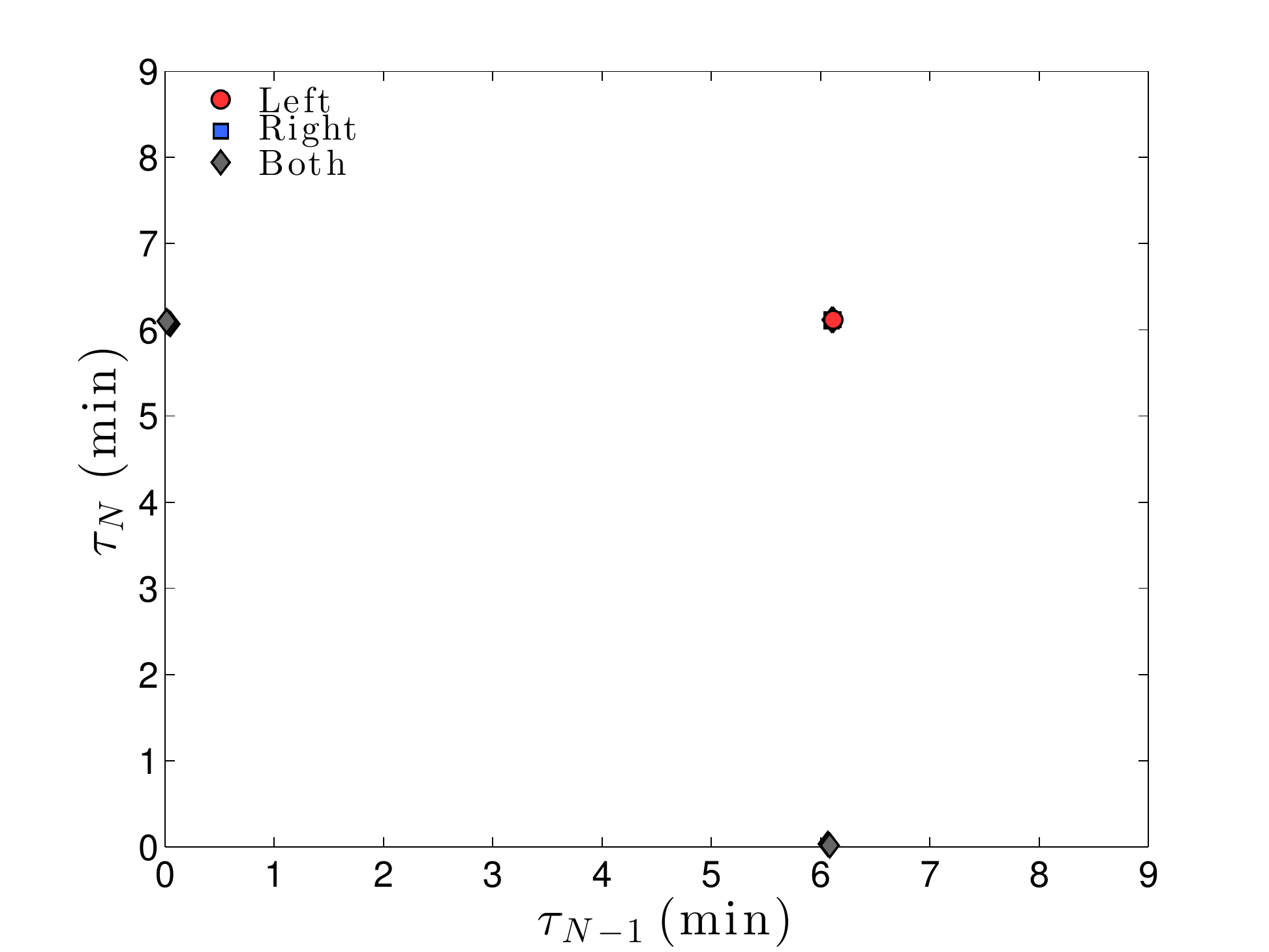}
 	\end{minipage}%
 	\caption{Correlation between the time between eruptions of the event denoted $N$ and the previous event denoted $N-1$. (a) $\alpha=0.2$. (b) $\alpha=0.7$. (b) $\alpha=0.9$.}
		\label{fig:double_model}
\end{figure}

In order to rationalize the previous observations, we develop a theoretical modeling of an interconnected geyser. The model is based on the idea that when a reservoir reaches its boiling temperature and erupts, it empties partially the vent and reduces the height of the water column for the other reservoir. Thus, depending on the temperature of the second reservoir, the eruption of the first reservoir may entrain the second one. In the model, we assume that the temperature of each reservoir follows a perfect saw-tooth evolution which is governed successively by the equations: $dT_r/dt= M \mathcal{P}  / c$ and $dT_r/dt= - M \mathcal{P}_d / c$ in order to stay in the range of temperature $T_{r,i} < T_r < T_b (H)$. We consider that the evolution of the temperature in both reservoirs are independent unless the temperature of one reservoir is larger than a temperature threshold $T_t = T_b ((1- \alpha) H)$ when the other erupts. In such case, the eruption of a reservoir provokes the eruption of the other one. This approach provides a prediction of times between eruptions for each reservoir. Figure \ref{fig:double_model} shows the correlation plot of the times between eruptions for the three values of the coupling parameters inspected experimentally in paragraph \ref{sec:double_exp}.

One notices in Fig. \ref{fig:double_model} that despite the large spreading of the experimental data is not reproduced by the model, the general trends are similar to the ones observed in section \ref{sec:double_exp}. Indeed, the times between eruptions of the total system occupy a trapezoidal domain which vanishes at large coupling parameters $\alpha$. Moreover, the spreading of the times between eruptions for each reservoir increases with $\alpha$ up to a critical value where the two reservoirs act as a single one.

Besides, the dispersion observed in Fig. \ref{fig:double_exp2} for $\tau$ can be approached by the theory developed in Section \ref{sec:model} for a toy geyser with a single reservoir. Indeed, Eq. (\ref{eq:geyser_period}) and (\ref{eq:geyser_tau_e}) yield the time between two eruptions of a reservoir if the other one does not experience any eruption. The substitution in Eq. (\ref{eq:geyser_period}) and (\ref{eq:geyser_tau_e}) of the boiling temperature $T_b(H)$ by its value $T_b ((1- \alpha) H)$ when the other reservoir has erupted, provides the minimal time between eruptions that a reservoir can follow. The difference between these two extreme cases provides an estimation of the dispersion observed for $\tau$. We delimited in Fig. \ref{fig:double_exp2} the domain in which the time between eruptions is supposed to be confined according this approach. The discrepancy observed between the experimental dispersions and the theoretical ones may come from the variability of boiling processes.

\section*{Conclusions}

The study of a toy geyser reveals the successive operations which lead to periodic eruptions of steam and hot water. In a geyser, a geometrical constriction prevents thermal convection of water from a heated reservoir toward a temperated pool. Thus, the water in the reservoir slowly accumulates heat energy by being superheated relatively to the pool level. Such a situation holds till the water in the reservoir boils and the vaporization ejects the water in the vent, a phenomena which generates a large release of energy in a short period of time. The slow accumulation of energy and its fast release is characteristic of catastrophic natural events such as earthquakes, volcanic eruptions or thunderbolts. We showed with the example of a toy geyser that the conditions to get such an accumulation of thermal energy and a periodic release are strict. This may explain why geysers are rare objects among other geothermal sources such as fumaroles, hot springs and boiling springs. Moreover, toy models of geyser allow us to understand their sensitivity towards external parameters such as water inflow, barometric pressure or pool temperature. 

However, a major difference exists between toy geysers and natural ones: the latter generally adopt a complex dynamics which strongly differs from the clock-like precision of toy geysers. The complexity of real geysers has been approached with a toy geyser having two interconnected reservoirs. Such a set-up showed a complex time behavior which has been studied statistically and rationalized theoretically. The rareness of faithful geysers in nature reveals that very few of them have a single and independent reservoir. Interconnected and complex plumbing systems seem to prevail in nature leaving room for the complexity rather than simplicity.

Besides, the present study lets several open questions such as the deep understanding of the phenomenon of bubble clogging, the geyser operation for large vent radius or the exact description of heat transfer during a boiling process in a superheated liquid. Also, this work opens new perspectives of research such as geysers made with a liquid differing from water or with a liquid ables to let solid deposition in order to study the stability of the plumbing systems over time.

\section*{Acknowledgments}

This research has been funded by the Inter-university Attraction Poles Programme (IAP 7/38 MicroMAST) initiated by the Belgian Science Policy Office. SD thanks the FNRS for financial supports. MB acknowledges support as a FNRS-FRIA Fellow. We are grateful to Alice Dubus who realised the glass reservoirs we dreamed about. Tadd Truscott and Zhao Pan are thanked for their fascinations concerning this subject and the excellent idea of interconnecting two reservoirs. The authors acknowledge Herv\'e Caps, Christophe Clanet, Laurent Maquet, Julien Schokmel, Alexis Duchesne for eruption of ideas and precious comments concerning this work. 
\bibliographystyle{unsrt}
%\bibliography{biblio}

\end{document}